\documentclass[aip,jcp,reprint]{revtex4-1}
\usepackage[colorlinks]{hyperref}
\usepackage[T1]{fontenc}
\usepackage{amsmath,amssymb}
\usepackage{braket}
\usepackage{units}
\usepackage{graphicx}
\usepackage[para]{threeparttable}
\usepackage{booktabs}
\usepackage{dcolumn}
\usepackage{enumitem}
\setlist{nosep}

\newcolumntype{.}{D{.}{.}{-1}}

\draft 

\usepackage[colorinlistoftodos,prependcaption,textsize=tiny]{todonotes} 
\usepackage{color} 
\newcommand{\angstrom}{\mbox{\normalfont\AA}}
\newcommand{\tnu}{\tilde{\nu}}
\newcommand{\subst}[1]{\textbf{\textsf{#1}}}

\newcommand{\PD}{\textbf{\textsf{PeDMA}}}
\newcommand{\Pe}{\textbf{\textsf{Pe}}}
\newcommand{\DMA}{\textbf{\textsf{DMA}}}
\newcommand{\C}{\textbf{\textsf{C153}}}

\newcommand{\be}{\begin{equation}}
\newcommand{\ee}{\end{equation}}
\newcommand{\nocontentsline}[3]{}
\newcommand{\tocless}[2]{\bgroup\let\addcontentsline=\nocontentsline#1{#2}\egroup}

\begin{document}


\title{How good is the generalized Langevin equation to describe the dynamics of photo-induced electron transfer in fluid solution?} 



\author{Gonzalo Angulo}
\email[]{gangulo@ichf.edu.pl}
\affiliation{Institute of Physical Chemistry, Polish Academy of Sciences, Kasprzaka 44/52, 01-224 Warsaw, Poland}

\author{Jakub J\k{e}drak}
\affiliation{Institute of Physical Chemistry, Polish Academy of Sciences, Kasprzaka 44/52, 01-224 Warsaw, Poland}

\author{Anna Ochab-Marcinek}
\affiliation{Institute of Physical Chemistry, Polish Academy of Sciences, Kasprzaka 44/52, 01-224 Warsaw, Poland}

\author{Pakorn Pasitsuparoad}
\affiliation{Institute of Physical Chemistry, Polish Academy of Sciences, Kasprzaka 44/52, 01-224 Warsaw, Poland}

\author{Czes\l aw Radzewicz}
\affiliation{Institute of Physical Chemistry, Polish Academy of Sciences, Kasprzaka 44/52, 01-224 Warsaw, Poland}

\author{Pawe\l\ Wnuk}
\affiliation{Max-Planck Institute of Quantum Optics, Ludwig Maximilian University of Munich, 85748 Garching, Germany}

\author{Arnulf Rosspeintner}
\email[]{arnulf.rosspeintner@unige.ch}
\affiliation{Department of Physical Chemistry, University of Geneva, Quai Ernest-Ansermet 30, CH-1211 Geneva, Switzerland}


\date{\today}

\begin{abstract}
The dynamics of unimolecular photo-triggered reactions can be strongly affected by the surrounding medium for which a large number of theoretical descriptions have been used in the past. An accurate description of these reactions requires knowing the potential energy surface and the friction felt by the reactants. Most of these theories start from the Langevin equation to derive the dynamics, but there are few examples comparing it with experiments. Here we explore the applicability of a Generalized Langevin Equation (GLE) with an arbitrary potential and a non-markovian friction. To this end we have performed broadband fluorescence measurements with sub-picosecond time resolution of a covalently linked organic electron donor-acceptor system in solvents of changing viscosity and dielectric permittivity. In order to establish the free energy surface (FES) of the reaction we resort to stationary electronic spectroscopy. On the other hand, the dynamics of a non-reacting substance, Coumarin~153, provide the calibrating tool for the non-markovian friction over the FES, which is assumed to be solute independent. A simpler and computationally faster approach uses the Generalized Smoluchowski Equation (GSE), which can be derived from the GLE for pure harmonic potentials. Both approaches reproduce the measurements in most of the solvents reasonably well. At long times, some differences arise from the errors inherited from the analysis of the stationary solvatochromism and at short times from the excess excitation energy. However, whenever the dynamics become slow the GSE shows larger deviations than the GLE, the results of which always agree qualitatively with the measured dynamics, regardless of the solvent viscosity or dielectric properties. The here applied method can be used to predict the dynamics of any other reacting system, given the FES parameters and solvent dynamics are provided. Thus no fitting parameters enter the GLE simulations, within the applicability limits found for the model in this work.
\end{abstract}


\maketitle 

\tableofcontents
\makeatletter
\let\toc@pre\relax
\let\toc@post\relax
\makeatother 

\section{Introduction}
For many chemical reactions a classical kinetics description using rate constants is sufficient, at least for their description at long times since their start. The Arrhenius empirical picture\cite{arrhenius_ZPC_1889} and the Transition State Theory,\cite{wigner_ZPCAB_1932, eyring_JCP_1935, chandler_JCP_1978} still practical and widely used, imply and assume an energetic barrier between reactants and products, which the reacting system has to surpass for the reaction to take place. This step determines the reaction rate constant. However, in liquids the passage over the activation barrier is not the only part to be taken into account.\cite{laidler_1987} This is especially the case for reactions with a large coupling between the educts and the products, or, in other terms, for reactions that proceed without jumps between different potential energy surfaces, i.e.\ adiabatic reactions.\cite{nitzan_2006} The much debated Kramers' theory (extended later by Pollack and others\cite{pollak_JCP_1986, *melnikov_JCP_1986, *pollak_JCP_1989, pollak_C_2005}) was one of the first to try to explain how the energy-changes of the reacting species in the bottom of the potential energy surface of the reactants couple to the movement over the barrier.\cite{kramers_P_1940} Such a movement can not directly be linked to a molecular translation or vibration but to a journey over different configurations of the reacting species with the surrounding medium. Obviously, one of the most complex questions concerns the nature of the path that dominates this journey - the associated reaction coordinate(s).\cite{pollak_C_2005} Eventually, in most of the cases all relevant degrees of freedom are projected into a general reaction coordinate.\cite{zwan_JCP_1983, pollak_CPL_1986, kawai_PCCP_2011} In the case of electron or charge transfer reactions - assuming no other degrees of freedom are relevant or fast enough - this coordinate is the polarization of the solute-solvent complex.\cite{zusman_CP_1980, barzykin_ACP_2002} Under these conditions dynamical solvent effects may become relevant. The Marcus-theory for electron transfer reactions has accommodated these effects either through the direct introduction of the dielectric relaxation time into the rate constant,\cite{zusman_CP_1980} or by considering the diffusion problem over the potential- or free-energy surface (PES, FES) with sudden jumps from the reactants' to the products' well.\cite{sumi_JCP_1986a} The latter is already a dynamic theory for the reaction-event based on the Smoluchowski diffusion operator for one dimension in the presence of a potential. An alternative to these models consists in solving the time dependent Schr\"{o}dinger equation introducing the friction introduced by the solvent in the Hamiltonian, which has the advantage of being potentially fully ab-initio at the cost of long computational time for each specific case.\cite{thallmair_JPCL_2014}

Most of the above-referred treatments try to obtain analytical formulas for the rate constant, as for example the powerful Grote-Hynes theory,\cite{grote_JCP_1980} or at least a reaction-coordinate dependent rate coefficient. In the core of these theories lies the generalized Langevin equation (GLE), which is valid for any potential shape and for any kind of friction - with or without memory. Under certain circumstances, like a harmonic potential, an equivalent generalized Fokker-Planck (FP) equation (an equation of diffusion in phase space) can be derived.\cite{nitzan_2006} Moreover, with a single coordinate in the projected phase space this FP expressions can be rewritten as a Generalized Smoluchowski equation (GSE) without losing details on the friction, as the diffusion coefficient is time-dependent.\cite{gudowska_APPB_1995} Further simplified expressions can be derived if the system is overdamped, meaning that the second derivative of the reaction coordinate with respect to time (i.e.\ the ``acceleration'') can be neglected, and the diffusion coefficient in the Smoluchowski equation becomes a constant. Such a hierarchy of equations suggests that the most assumption-free and therefore most general approach would be using directly the GLE, instead of a Smoluchowski approach or an analytical expression for the rate constant. In doing so, the details of the reaction or in general the dynamics, would not be lost due to the approximations. For example, in solvation dynamics it is often observed that at short times the response of the solvent to a perturbation like the change of the solute dipole moment after light absorption is not diffusive but impulsive.\cite{[{}][{ and refs. 28, 38-43 therein.}]karunakaran_JPCA_2008} This detail is lost in the simpler of the Smoluchowski equations but is conserved in the GLE and in the GSE. The disadvantage of the GSE resides in the fact, that it is only strictly valid for harmonic potentials, although it was extended to cases with parabolic barriers by Okuyama and Oxtoby\cite{okuyama_JCP_1986} and has been used in various occasions with other potentials.\cite{tominaga_JPC_1991, tominaga_JPC_1991a, meer_JCP_2000, *heisler_JPCB_2009, *kondo_JPCB_2009}

Surprisingly, there are few works in the literature in which measured electron transfer dynamics are compared to the output of GLE simulations.\cite{smith_1992, cherepanov_BJ_2001, min_PRL_2005} The most noticeable examples are those of the groups of Barbara and Fonseca in the late 80s and early 90s.\cite{kang_CP_1990, tominaga_JPC_1991a, tominaga_JPC_1991} This renders the GLE, as applied to ultrafast chemical reactions in solution, still a hypothesis rather than a contrasted theory.

The difficulty in computing the GLE lies in its stochastic form, with a random variable and a noise term with time as the ordering parameter,\cite{nitzan_2006} which requires performing numerous simulation trajectories to obtain meaningful statistics. The ingredients needed to perform these simulations are the PES and the friction felt by the reacting system. It can be formulated the following way:\cite{grote_JCP_1980}
\begin{equation}
\ddot{z} = -\dfrac{1}{m_{\rm L}}\dfrac{\partial V(z)}{\partial z} - \int_0^t \eta(t - \tau) \dot{z} {\rm d}\tau + R(t),
\label{eq:gle}
\end{equation}
where $\eta(t)$ denotes the friction term and $R(t)$ is the noise term. The random variable, $z$, can be normalized and takes the value 0 for the reactants' minimum and 1 for the products minimum, in the potential energy surface $V$. The mass, $m_{\rm L}$, is associated to the ``heaviness'' of the polarization of the medium and can be extracted from the rotational motion of the individual solvent molecules.\cite{hynes_JPC_1986} The integral term stands for the non-markovian friction, $\eta(t)$, against the movement of the system-associated particle. Beware, that in the present context this friction is not purely mechanical, but reflects the resistance of the dielectric to a change in the electric field associated to the charge redistribution in the solute, as for example those occurring during an electron transfer. This memory can be understood as an extended duration of the correlation function of the medium beyond the delta-function response that characterizes pure Debye-solvents.\cite{debye_1929} This friction has the same origin as the noise, which are the configuration fluctuations of the solute-solvent entity. Therefore, they are related through the second fluctuation-dissipation theorem:\cite{kubo_RPP_1966}
\begin{equation}
\braket{R(t)R(t+t')} = \eta(t') \braket{\dot{z}^2}.
\end{equation}
Determining the PES (which in condensed phases is actually a FES)\cite{chandler_JCP_1978} in the excited state is not an easy task from QM calculations. Thus having experimental data which can elucidate its shape is extremely advantageous. The friction, in the case of charge transfer reactions is solely modulated by the solvent polarization and not by internal modes of the solute. It can be calculated from the dielectric relaxation spectrum of the solvent.\cite{sajadi_JPCA_2009} Unfortunately, we do not count on many experiments of this sort yet. Another approach would be to somehow extract it from a reference measurement for which the FES is well known.

\begin{figure}[!tp]
	\includegraphics{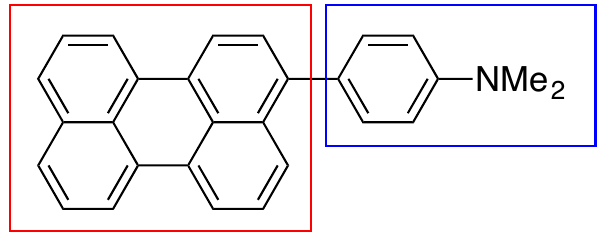}
\caption{Chemical structure of the donor (\DMA, blue box) -acceptor (\Pe, red box) molecule (\PD) under investigation.}
\label{fig:PeDMA}
\end{figure}

This is the spirit of the current work: we have tried to obtain both of the above mentioned quantities from auxiliary measurements in order to explain the solvent dependent dynamics of a charge transfer reaction in a model compound (cf.\ Figure \ref{fig:PeDMA}), perylene (\Pe) covalently linked to the donor N,N-dimethylaniline (\DMA).\cite{banerji_JPCA_2008} The dynamics have been obtained by means of fluorescence up-conversion spectroscopy (FLUPS).\cite{schanz_APL_2001, *zhang_RSI_2011, *sajadi_APL_2013, *gerecke_RSI_2016} This method has a significant advantage over other conventional techniques tracking fast fluorescent events in the sub-ps to the ns timescales. FLUPS directly provides photometrically correct emission spectra and does not require - useful and correct but cumbersome and error-prone - spectral reconstruction methods.\cite{maroncelli_JCP_1987} Thus the dynamics of the evolution over the excited state FES are reliably monitored. In order to obtain the parameters needed to calculate the FES, stationary absorption and fluorescence measurements of \PD\ have been performed in a relatively large set of solvents covering a wide range of dielectric properties. The application of a continuum solvation model\cite{liptay_ZNAPS_1965, rosspeintner_C_2010} to explain these data leads to dipole moments and energies that are used to calculate the FES. For the friction, we have performed FLUPS measurements of \C\ in the very same solvents as for \PD. \C\ has been extensively studied and is assumed to behave as if its FES in the excited state is purely harmonic.\cite{maroncelli_JCP_1987, maroncelli_JML_1993, horng_JPC_1995} This approximation allows us to reconstruct the friction that is later employed in the GLE to simulate the polarization changes during the reaction of charge transfer within \PD.

In addition to the GLE simulations we have also calculated the purely diffusive motion over the FES using the Smoluchowski operator with a time-dependent diffusion coefficient obtained from the \C\ measurements.\cite{hynes_JPC_1986} The resulting GSE equation is given by
\begin{equation}
\dfrac{\partial \rho(z,t)}{\partial t} = D(t) \dfrac{\partial}{\partial z} \left[\dfrac{\partial}{\partial z} + \frac{1}{k_{\rm B}T}\frac{\partial}{\partial z}F(z)\right] \rho(z,t).
\label{eq:GSE}
\end{equation}
It describes the temporal behaviour of the population distribution $\rho(z,t)$ along an arbitrary FES (or PES), $F(z)$. The solution is strictly valid only for harmonic potentials, but in our case the large coupling in \PD\ leads to a form very close to that and therefore very close to gaussian distributions of the population over S$_1$.

The article is organized as follows: the next section contains the details of the methods and materials used in the experiments. Then the spectroscopic measurements are presented, first the stationary with a discussion of the observations, followed by the time-resolved observations including broadband fluorescence and transient absorption. The next chapter describes how the FES are obtained from the stationary measurements with a discussion of the parametrization. Then the procedure to obtain the dynamic quantities for the GLE and the GSE is presented, as well as the details of the calculation procedures with these models. Finally, we compare the simulations with the experimental data and discuss the results.

The comparison between the two models and the experiments leads us to conclude that the GLE is superior in reproducing the experiments for the present case. An interesting observation is that the dynamics, despite being often slower than the simple Debye relaxation times for some of the solvents, are still fully controlled by the solvent relaxation dynamics. This is a result of the interplay between the shape of the FES and the friction, as described by the GLE.

\section{Methods and Materials}
\subsection{Chemicals}
The synthesis of 3-(\emph{p}-N,N\--dimethyl\-amino\-phenyl)\-perylene (\subst{\PD}) is described in ref.\,\citenum{banerji_JPCA_2008}. Coumarin~153 (\C, CAS  53518-18-6) and BBOT (2,5-bis[5-tert-butylbenzoxazolyl(2)]thiophene, CAS 7128-64-5) were used as received. The solvents were of the highest commercially available purity and were used without further purification. In addition to the pure solvents we used two binary solvent mixtures: Benzyl acetate / dimethylsulfoxide allows to vary the dielectric constant without significantly changing neither the refractive nor the viscosity.\cite{rosspeintner_2008} Dimethylsulfoxide / glycerol allows for changing the solvent viscosity without changing neither the dielectric constant nor the refractive index.\cite{angulo_PCCP_2016}

\subsection{Spectroscopy}
Steady-state absorption and emission spectra were recorded on a Cary 50 and a FluoroMax-4 (bandpass of \unit[1]{nm}), respectively. The wavelength sensitivity of the fluorimeter has  been determined using a set of secondary emissive standards.\cite{gardecki_AS_1998} Fluorescence quantum yields were determined according to 
\begin{equation}
\phi_{\rm f} = \phi_{\rm r} \left(\dfrac{n_{\rm s}^2}{n_{\rm r}^2} \right) \left(\dfrac{I_{\rm s}}{I_{\rm r}} \right) \left(\dfrac{A_{\rm r} \cdot 10^{-d_{\rm eff}\cdot A_{\rm r}}}{A_{\rm s} \cdot 10^{-d_{\rm eff}\cdot A_{\rm s}}} \right) 
\end{equation}
where $n_x$ is the solvent refractive index, $I_x$ is the integrated emission intensity, $A_x$ is the absorbance of sample $x$ (either \underline{s}ample or \underline{r}eference) at the excitation wavelength and $d_{\rm eff}$ is the effective distance from the cuvette excitation window to the point at which the luminescence is observed and which has been determined experimentally as proposed in reference \citenum{kubista_TA_1994}. The $\phi_{\rm r}$ values of \C\ in methanol and tetrahydrofuran were used as references.\cite{Lewis_CPL_1998} Samples for measuring fluorescence lifetimes and quantum yields were contained in septa-sealed quartz cuvettes (Starna, 3/GL14/Q/10) with \unit[10]{mm} pathlength and were degassed with argon prior to measurement during \unit[10]{mins}.

Transition dipole moments have been calculated according to Birks\cite{birks_1970}
\begin{subequations}
\label{eq:M}
\begin{align}
\mu_{\rm a} & = 9.584\cdot 10^{-2} \sqrt{\dfrac{1}{n} \int_{S_1} \dfrac{\epsilon(\tnu)}{\tnu}{\rm d}\tnu} \label{eq:Ma}\\
\mu_{\rm f} & = 1.7857\cdot10^3 \sqrt{\dfrac{k_{\rm rad}}{n^3} \dfrac{\int I(\tnu)\tnu^{-3}{\rm d}\tnu}{\int I(\tnu){\rm d}\tnu}},
\label{eq:Mf}
\end{align}
\end{subequations}
where $n$ denotes the solvent refractive index and $k_{\rm rad}$ is the radiative rate of the excited state, calculated according to $k_{\rm rad} = \phi_{\rm f}/\tau_{\rm f}$, and $\tau_{\rm f}$ is the fluorescence lifetime. $\epsilon(\tnu)$ and $I(\tnu)$ are the extinction coefficient spectrum, in \unitfrac[]{L}{mol cm}, and the fluorescence spectrum versus $\tnu$, in cm$^{-1}$, respectively. 

Nanosecond, single-wavelength, time-resolved fluorescence experiments were performed using a home-built time-correlated single-photon-counting set-up described in the SI of ref.\ \citenum{rosspeintner_JACS_2014}. In brief, the sample was excited at \unit[400]{nm} using a pulsed laser diode with an approximate pulse duration of \unit[50]{ps}. The time-resolution of these experiments, as judged from the full-width half maximum of the instrument response function (IRF), measured with a dilute scattering solution of ludox in water, amounts to approximately \unit[200]{ps}.

Femtosecond time-resolved broadband fluorescence up-conversion spectra (FLUPS) were measured on an apparatus identical to the one described in ref.\ \citenum{gerecke_RSI_2016}, with an approximate time-resolution of \unit[170]{fs} (fwhm of the instrument response function). The experimental spectra were corrected for the wavelength dependent detection sensitivity using a set of secondary emissive standards (covering the range from \unit[415-720]{nm}) and the temporal chirp, using the wavelength-independent instantaneous response of BBOT in the solvents under investigation as reference.\cite{zhang_RSI_2011}

Femtosecond time-resolved broadband transient absorption spectra were measured on an apparatus described in ref.\,\citenum{lang_RSI_2013}. In brief, samples were excited at \unit[400]{nm} with \unit[100]{fs} pulses, while broadband probing in the range from \unit[370]{nm} to \unit[720]{nm} was achieved using a white-light continuum, generated in a \unit[3]{mm} thick CaF$_2$.

The samples were placed in commercial quartz cells with an optical pathlength of \unit[1]{mm} (Starna, 1/GS/Q/1). Bubbling with nitrogen during the measurement ensured the absence of traces of oxygen as well as a sufficient sample-exchange to avoid photodecomposition. All measurements were performed at magic angle conditions, except where noted, by setting the appropriate polarization of the pump-pulse with respect to the white light continuum (TA) or the gate pulse (FLUPS). Sample concentrations were kept below \unitfrac[10$^{-4}$]{mol}{L} and pump fluences per pulse were below \unitfrac[4]{mJ}{cm$^2$}.

All ultrafast measurements were performed at \unit[$20\pm1$]{$^\circ$C}.

DFT calculations on the B3LYP level of theory with the  6-31G-(d,p) basis set were performed using Gaussian~09.\cite{Gaussian_2009} Solvent dependent permanent electric dipole moments were obtained from PCM calculations using the CPCM polarizable conductor calculation model.\cite{barone_JPCA_1998, cossi_JCC_2003}


\section{Experimental Results}
\subsection{Steady-State Spectroscopy}

\begin{figure}
\centering
	\includegraphics[]{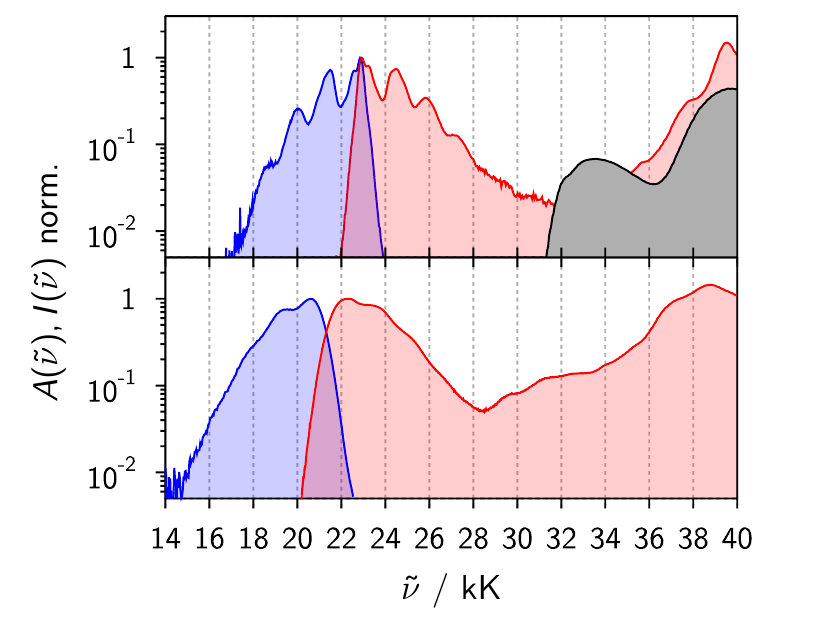}
\caption{Comparison of the absorption (red) and emission (blue) spectra of \Pe\ (upper panel) and \PD\ (lower panel) in cyclohexane. The upper panel also depicts the absorption spectrum of \DMA\ (grey) in cyclohexane, scaled according to the extinction coefficient with respect to \Pe\ (value taken from ref.\ \citenum{UVatlas_1967}). \unit[1]{kK}=\unit[1000]{cm$^{-1}$}.}
\label{fig:Pe_PeDMA}
\end{figure}

Figure~\ref{fig:Pe_PeDMA} compares the absorption and emission spectra of \PD\ with those of its two building blocks, \Pe\ and \DMA. The \PD-spectra are slightly red-shifted by approximately \unit[1500]{cm$^{-1}$} with respect to \Pe, but maintain the reasonable mirror-symmetry, which characterizes the \Pe-spectra (see Figure~S3 for transition dipole moment representation of the data from Figure\,\ref{fig:Pe_PeDMA}). Similarly, the local \DMA\ transitions also seem to be slightly red-shifted in \PD. While the vibration of approx.\ \unit[1500]{cm$^{-1}$}, which is responsible for the prominent vibronic progression in \Pe\ is preserved in \PD, its bandshape is significantly broader than that of \Pe. We tentatively attribute this to the lowering of symmetry in \PD\ and the ensuing increase in the number of vibrations capable of coupling to the electronic transition. This ``intrinsic'' broadening of the lineshape function of \PD\ translates into an increase of the observed Stokes shift\footnote{Usually defined as the energy difference between the lowest energy absorption maximum and the emission maximum.} in apolar cyclohexane, going from \unit[50]{cm$^{-1}$} for \Pe\ to approximately \unit[1760]{cm$^{-1}$} for \PD.

\begin{figure}[!htp]
\centering
	\includegraphics[]{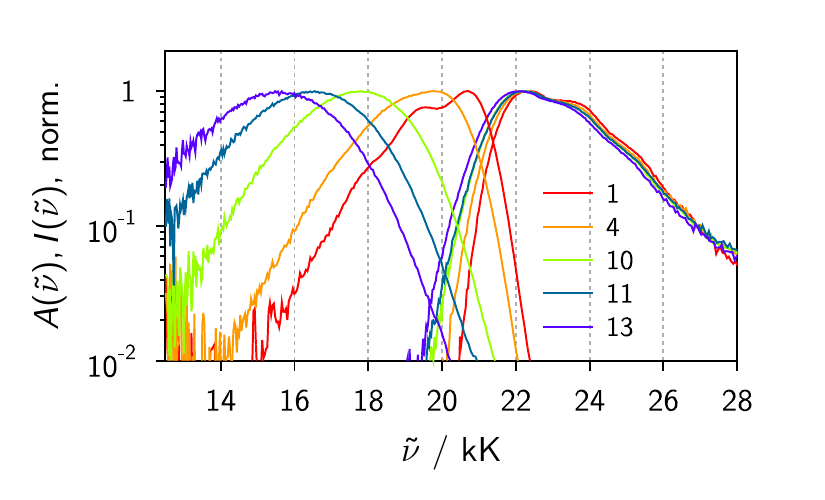}
\caption{Normalized absorption and fluorescence spectra of \PD\ in 5 selected solvents (see Table\,\ref{tab:exps} for solvent indices).}
\label{fig:solvato}
\end{figure}

The most striking difference between \PD\ and its building-block fluorophore, \Pe, is manifested when changing the solvent polarity (cf.\ Figure~\ref{fig:solvato}). On the one hand, \Pe\ shows almost negligible solvatochromism in absorption and emission, which can be relatively well correlated with changes in the solvent refractive index. \PD, on the other hand, shows a very pronounced bathochromic solvatofluorochromism of up to almost \unit[5000]{cm$^{-1}$}, which is accompanied by a total loss of vibronic structure, while the absorption spectra - in analogy to \Pe\ - show neither a dramatic change in bandshape nor position.

For further analysis we calculated the first and second moments of the absorption and emission spectra, which are defined as follows\cite{spanos_1999, braem_PCCP_2012}
\begin{subequations}
\label{eq:mom} 
\begin{align}
m_1 & = \int \tnu S(\tnu) {\rm d}\tnu,\\
m_2 & = \int (\tnu - m_1)^2 S(\tnu) {\rm d}\tnu,
\end{align}
\end{subequations}
where $S(\tnu)$ is the area normalized spectrum of interest. $m_1$ corresponds to the average spectral position, while $m_2$ represents the spectral variance. The values in all solvents are summarized in Table~\ref{tab:exps} together with other relevant steady-state parameters. The emission spectra in polar solvents partially expanded beyond the detection-wavelength range of the fluorimeter (see e.g.\ emission spectra in solvents 11 and 13 in Figure~\ref{fig:solvato}). Thus, rather than extracting the moments directly from the corresponding emission spectra we fitted them with the convolution of a spectral lineshape function, $L_x(\tnu)$, accounting for the vibronic structure and intrinsic linebroadening, with a gaussian distribution, $\rho(\delta)$, which accounts for inhomogeneous broadening due to the solvent environment.\footnote{Here we do not account for broadening due to electronic dephasing.}\cite{horng_JPC_1995}
\begin{subequations}
\label{eq:conv} 
\begin{align}
A(\tnu) & \propto \tnu \int \rho(\delta) L_{\rm a}(\tnu - \delta) {\rm d}\delta\\
I(\tnu) & \propto \tnu^3 \int \rho(\delta) L_{\rm f}(\tnu - \delta) {\rm d}\delta,
\end{align}
\end{subequations}
where the gaussian distribution is given by 
\begin{equation}
\rho(\delta) = \frac{1}{\sigma\sqrt{2\pi}} \exp\left[- \frac{(\delta - \delta_0)^2}{2\sigma^2} \right]. \label{eq:distroG}
\end{equation}
Here we used the oscillator distributions of the absorption and emission spectrum of \PD\ in the apolar solvent {\it n}-hexane as lineshape function, $L_{x}(\tnu)$. 
\begin{subequations}
\label{eq:L}
\begin{align}
L_{\rm a}(\tnu) & \propto \tnu^{-1} A(\tnu) \\
L_{\rm f}(\tnu) & \propto \tnu^{-3} I(\tnu)
\end{align}
\end{subequations}
Exemplary fits of these equations to the experimental spectra are shown in Fig.~S4.

The absorption and emission transition dipole moments of \PD\ are identical in cyclohexane (\unit[6.2]{D}) and by almost 30\% larger than the ones of \Pe\ (\unit[4.8]{D}). While in apolar solvents the transition dipole moments for absorption and emission are identical to within experimental error, increasing the solvent polarity leads to a slight decrease of the emission transition dipole moments in \PD\ (\emph{vide infra}, Figure~\ref{fig:ss_fit}).

\begin{table*}
	\begin{threeparttable}[c]
			\caption{Solvent properties (at \unit[20]{$^\circ$C}) and spectral characteristics of \PD.\tnote{a}}
	\begin{ruledtabular}
	\begin{tabular}{lr...........}
		 &  & \multicolumn{2}{c}{solvent properties} & \multicolumn{4}{c}{spectral characteristics} & & \\
			\cline{3-4} \cline{5-9}
			\rule{0pt}{3ex}
			solvent & no. & \multicolumn{1}{c}{$\epsilon$} & \multicolumn{1}{c}{$n$} & \multicolumn{1}{c}{$m_{\rm 1a}$} & \multicolumn{1}{c}{$\sqrt{m_{\rm 2a}}$} & \multicolumn{1}{c}{$m_{\rm 1f}$} & \multicolumn{1}{c}{$\sqrt{m_{\rm 2f}}$} & \multicolumn{1}{c}{$\Gamma_{\rm f}$} & \multicolumn{1}{c}{$\mu_{\rm a}$} & \multicolumn{1}{c}{$\mu_{\rm f}$} & \multicolumn{1}{c}{$\tau_{\rm f}$} & \multicolumn{1}{c}{$\phi_{\rm f}$} \\
				& 	&	&	& \multicolumn{5}{c}{(kK)} & \multicolumn{1}{c}{(D)} & \multicolumn{1}{c}{(D)} & \multicolumn{1}{c}{(ns)} &		\\
		\cline{1-13}
					$n$-hexane  & 1 & 1.88 & 1.375 & 23.19 & 1.20 & 19.75 & 1.26 & 0.00 & 6.19 & 6.23 & 3.51 & 0.82 \\ 
					$c$-hexane  & 2 & 2.02 & 1.426 & 23.13 & 1.22 & 19.66 & 1.27 & 0.21 & 5.98 & 6.21 & 3.44 & 0.87 \\ 
					$n$-pentyl ether  & 3 & 2.77 & 1.412 & 23.04 & 1.25 & 19.12 & 1.32 & 0.92 & 6.18 & 6.28 & 3.75 & 0.86 \\ 
					$n$-butyl ether  & 4 & 3.08 & 1.399 & 23.06 & 1.24 & 19.06 & 1.34 & 1.06 & 6.23 & 6.29 & 3.80 & 0.84 \\ 
					$i$-propyl ether  & 5 & 3.88 & 1.368 & 23.13 & 1.23 & 18.89 & 1.37 & 1.30 & 6.29 & 6.45 & 3.96 & 0.84 \\ 
					ethyl ether  & 6 & 4.20 & 1.352 & 23.15 & 1.25 & 18.67 & 1.40 & 1.52 & 6.26 & 6.36 & 4.20 & 0.81 \\ 
					chloroform  & 7 & 4.89 & 1.446 & 22.90 & 1.29 & 18.16 & 1.41 & 1.58 & 6.16 & 6.17 & 4.13 & 0.83 \\ 
					butyl acetate  & 8 & 5.01 & 1.395 & 23.02 & 1.26 & 17.94 & 1.49 & 1.99 & 6.31 & 6.29 & 4.56 & 0.83 \\ 
					ethyl acetate  & 9 & 6.02 & 1.372 & 23.07 & 1.26 & 17.66 & 1.57 & 2.26 & 6.31 & 6.27 & 4.92 & 0.81 \\ 
					tetrahydrofurane  & 10 & 7.58 & 1.407 & 22.94 & 1.29 & 17.42 & 1.53 & 2.17 & 6.27 & 6.16 & 4.83 & 0.80 \\ 
					butyronitrile  & 11 & 24.83 & 1.384 & 22.94 & 1.30 & 16.24 & 1.67 & 2.73 & 6.35 & 6.05 & 6.14 & 0.76 \\ 
					acetonitrile  & 12 & 35.94 & 1.344 & 23.03 & 1.28 & 15.55 & 1.74 & 2.98 & 6.45 & 5.87 & 6.19 & 0.59 \\ 
					dimethylformamide  & 13 & 36.71 & 1.430 & 22.80 & 1.33 & 15.35 & 1.73 & 2.93 & 6.25 & 5.74 & 5.01 & 0.52 \\ 
					dimethylsulfoxide  & 14 & 47.01 & 1.478 & 22.67 & 1.36 & 14.81 & 1.76 & 2.99 & 6.15 & 5.48 & 4.15 & 0.40 \\[1ex] 

					DB $x_{\rm D} = 0.10$  & 20 & 7.24 & 1.501 & 22.78 & 1.33 & 17.22 & 1.56 & 2.27 & - & 6.20 & 4.41 & 0.87 \\ 
					DB $x_{\rm D} = 0.16$  & 21 & 8.20 & 1.500 & 22.76 & 1.32 & 16.95 & 1.59 & 2.39 & - & 6.13 & 4.63 & 0.86 \\ 
					DB $x_{\rm D} = 0.25$  & 22 & 9.88 & 1.499 & 22.78 & 1.34 & 16.62 & 1.62 & 2.50 & - & 6.09 & 4.90 & 0.84 \\ 
					DB $x_{\rm D} = 0.33$  & 23 & 11.66 & 1.498 & 22.74 & 1.34 & 16.38 & 1.63 & 2.57 & - & 6.10 & 5.08 & 0.84 \\ 
					DB $x_{\rm D} = 0.42$  & 24 & 14.05 & 1.496 & 22.73 & 1.34 & 16.13 & 1.66 & 2.68 & - & 6.03 & 5.21 & 0.80 \\ 
					DB $x_{\rm D} = 0.51$  & 25 & 16.95 & 1.495 & 22.72 & 1.35 & 15.87 & 1.68 & 2.75 & - & 6.02 & 5.27 & 0.76 \\ 
					DB $x_{\rm D} = 0.59$  & 26 & 20.01 & 1.493 & 22.72 & 1.35 & 15.70 & 1.69 & 2.81 & - & 5.97 & 5.24 & 0.72 \\ 
					DB $x_{\rm D} = 0.66$  & 27 & 23.15 & 1.491 & 22.71 & 1.34 & 15.57 & 1.77 & 2.88 & - & 5.95 & 5.19 & 0.69 \\ 
					DB $x_{\rm D} = 0.78$  & 28 & 29.72 & 1.488 & 22.70 & 1.35 & 15.30 & 1.73 & 2.96 & - & 5.88 & 4.96 & 0.61 \\ 
					DB $x_{\rm D} = 0.89$  & 29 & 37.38 & 1.484 & 22.68 & 1.35 & 15.08 & 1.75 & 3.06 & - & 5.82 & 4.59 & 0.53 \\ [1ex]

					$n$-pentanol  & 30 & 14.27 & 1.411 & 23.03 & 1.28 & 17.65 & 1.58 & 2.38 & - & 5.85 & 4.77 & 0.76 \\ 
					$i$-propanol  & 31 & 20.64 & 1.378 & 23.12 & 1.24 & 17.46 & 1.62 & 2.54 & - & 6.12 & 5.09 & 0.78 \\ 
					ethanol  & 32 & 24.55 & 1.361 & 23.11 & 1.26 & 17.05 & 1.63 & 2.70 & 6.36 & 6.10 & 6.14 & 0.73 \\ 
					methanol  & 33 & 32.66 & 1.328 & 23.15 & 1.25 & 16.58 & 1.64 & 2.94 & 6.42 & 6.10 & 5.01 & 0.57 \\ 
					ethylene glycol  & 34 & 38.69 & 1.432 & 22.84 & 1.32 & 15.61 & 1.79 & 2.94 & - & 5.40 & 3.28 & 0.32
	\end{tabular}
	\end{ruledtabular}
	\begin{tablenotes}		
		\item [a] solvent properties were either taken directly from ref.\,\citenum{riddick_1986} or recalculated to \unit[20]{$^\circ$C} from data in ref.\,\citenum{marcus_1998}. Data for the refractive indices and dielectric constants of the binary mixtures were taken from ref.\,\citenum{rosspeintner_2008}. $\tau_{\rm f}$ is the fluorescence lifetime. $m_{ix}$ is the $i$th spectral moment of absorption (a) or fluorescence (f). The spectra moments for absorption and fluorescence were calculated in the wavenumber range from \unit[19-26]{kK} and \unit[8-22.5]{kK}, respectively. $\Gamma_{\rm f}$ denotes the full-width at half maximum of the gaussian accounting for inhomogeneous broadening due to polar solvation, i.e. $\Gamma = \sigma \sqrt{8\ln 2}$ (with $\sigma$ being defined in eq.\,\eqref{eq:distroG}). DB denotes solvent mixtures of dimethylsulfoxide / benzyl acetate with a dimethylsulfoxide molar fraction of $x_{\rm D}$. Absorption spectra and fluorescence lifetimes were recorded at room temperature, $22\pm2 ^\circ$C. Emission spectra were recorded at $20^\circ$C.
	\end{tablenotes}
			\label{tab:exps}
\end{threeparttable}	
\end{table*}

\subsection{Time-resolved broadband Fluorescence}

\subsubsection{PeDMA}
\begin{figure}[!htp]
\centering
	\includegraphics[]{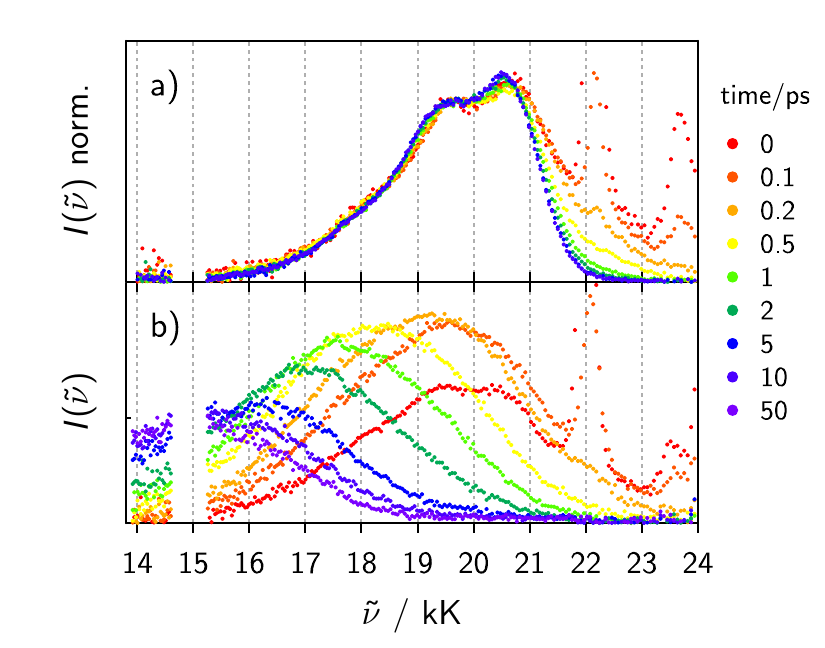}
\caption{a) Normalized FLUPS spectra of \PD\ in cyclohexane recorded under perpendicular excitation. b) FLUPS spectra of \PD\ in dimethylsulfoxide at selected time delays upon excitation at magic angle. Note that all dynamics are over after \unit[50]{ps}. The sharp peaks at approx.\ 22 and \unit[23.6]{kK} are caused by Raman scattering upon excitation at \unit[25]{kK}. We cut the data around \unit[15]{kK}, as they are strongly contaminated by the third harmonic of the gate.}
\label{fig:tr_spectra}
\end{figure}

Figure~\ref{fig:tr_spectra} shows representative FLUPS spectra of \PD\ in {\it c}-hexane and dimethylsulfoxide. In both examples the solvent Raman bands are clearly observable at short time-delays, when pump and gate still overlap temporally. In the apolar {\it c}-hexane a significant change in bandshape in the blue part of the spectrum is observed within the first picosecond. Immediately after excitation mirror symmetry is absent due to additional vibronic transitions at the high-energy side of the emission spectrum. Once these transitions have disappeared, while the red part of the spectrum virtually remains unaltered, mirror symmetry - as also observed in the steady-state spectrum in this solvent - is recovered. The additional initial vibronic transitions can be ascribed to emission from vibrationally excited states. In fact, exciting the sample with \unit[400]{nm} is expected to prepare the system in the $v_2$-state of the vibronic progression.\cite{kasajima_JPCA_2004} It is noteworthy, that no spectral narrowing of the red-part of the spectrum is observed, which is usually interpreted as a signature of vibrational cooling (cf.\ Figure\,\ref{fig:tr_spectra}a). We have performed measurements exciting the red edge of the absorption band at \unit[485]{nm} with the set-up described in ref.\,\citenum{angulo_PCCP_2016} and no significant differences could be found in the shape of the spectra in most of the wavelength range recorded, with respect to the spectra recorded upon excitation at \unit[400]{nm}. However, under these experimental conditions it is rather difficult to assess the short-time behavior of the spectra in the blue edge. In any case, the relaxation dynamics recorded upon excitation at \unit[400]{nm} are clearly influenced by intramolecular vibrational redistribution and vibrational cooling during the first picosecond. This is reflected in both the first and the second moments, as will be seen and discussed in the corresponding sections and is shown for \PD\ in {\it c}-hexane in Figure~S11.

In dimethylsulfoxide, a representative polar solvent, an initial emission spectrum with vibronic structure is observable within the IRF. However, it quickly evolves into a broad and unstructured band and with its first moment shifting by almost \unit[5]{kK} within \unit[50]{ps}.

Further analysis does not necessarily require using the full FLUPS data sets. Rather, we opted for determining the first and second moments of the data, $m_1(t)$ and $m_2(t)$ as defined by eq.\,\eqref{eq:mom}. This is more reasonable than simply following the band maximum since the spectral bandshape of \PD\ changes over time. In order to reduce the experimental noise, especially around the region of the doubled gate (ca.\,\unit[15]{kK}), we opted for fitting the convolution of an emission lineshape-spectrum and a gaussian of arbitrary position and width to the experimental spectra at each time-step and calculate the first two moments, $m_1(t)$ and $m_2(t)$, from these noise-free spectra (see eqs.\,\eqref{eq:conv}). The gaussian distributions were sufficiently close to the population distribution at any instance in time, in order to reproduce the experimental spectra, as judged from the weighted residuals.\footnote{We also tested the log-normal function as underlying distribution for convolution with the lineshape function, but did not obtain significantly improved fits. Thus we opted for the distribution with less degrees of freedom, i.e.\ the gaussian distribution.}

A popular way of representing spectro-temporal shifts consists in calculating a so-called spectral response function, $C_x(t)$,\cite{maroncelli_JCP_1987, horng_JPC_1995}
\begin{equation}
C_x(t) = \frac{x(t) - x(\infty)}{x(t_0) - x(\infty)},
\label{eq:Ct}
\end{equation}
where $x$ denotes either the first spectral moment, $m_1$, or the peak position of the band, $\tnu_{\rm p}$. $t_0$ is ideally chosen large enough to avoid problems with the finite duration of the instrument response function. However, if there are any discrepancies in the position of the spectra at short or long times between the spectrotemporal model (and the corresponding simulations) and the experimental results, the normalization procedure in eq.\,\eqref{eq:Ct} will significantly and artificially deform the dynamics and lead to potential misinterpretation. Thus, rather than performing any kind of transformation of the experimental observables via normalization, we opted to directly compare experimental and simulated moments.

\subsubsection{C153}

\begin{figure}[!htp]
\centering
	\includegraphics[]{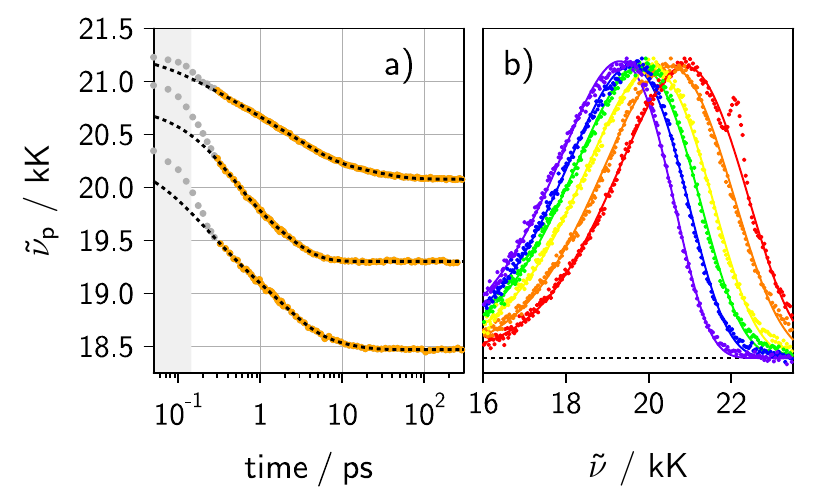}
\caption{a) Representative time dependences of the peak position of \C\ in solvents 8, 11 and 14 (from top to bottom), $\tnu_{\rm p}(t)$, (grey and orange dots) and the corresponding fits (dashed lines) to the data at times larger than \unit[0.3]{ps} (orange points) with $\tau_{\rm s} = 100$\,fs. The grey area indicates $2\sigma$ of the IRF. b) Representative FLUPS spectra of \C\ in butyronitrile (12) at 0.1, 0.2, 0.5, 1, 2 and \unit[20]{ps} time delay (from right to left / red to violet) and the corresponding log-normal fits (solid lines). The sharp peak at approx.\ \unit[22]{kK} is due to Raman scattering from the solvent.}
\label{fig:Ct_C153}
\end{figure}

In order to access the solvation dynamics of the reference compound, \C, we fitted a single log-normal function (see e.g. ref.\,\citenum{maroncelli_JCP_1987} for a definition) to the FLUPS spectra at each time-delay (cf.\ Figure~\ref{fig:Ct_C153}b). A multiexponential fit to the time-dependence of the log-normal peak maxima, $\tnu_{\rm p}(t)$, which in the past has been assumed to represent a good measure of the solvation dynamics,\cite{maroncelli_JCP_1987, kumpulainen_PCCP_2017} should allow obtaining an analytical description of the data. In order to account for the limited time-resolution of the set-up we fit the data only for times longer than \unit[0.3]{ps} (see Figure~\ref{fig:Ct_C153}a). Nonetheless, we recognize that sometimes a significant part of the solvation dynamics occurs within this initial time-window. In order to account for these dynamics and - what's more important - to estimate it's relative contribution to the overall solvation dynamics we estimate the peak position of the spectrum at time zero, $\tnu_{\rm p}(0)$, using the approach outlined in ref.\,\citenum{horng_JPC_1995}. Combined with the position of the steady-state spectrum, $\tnu_{\rm p}(\infty)$, the theoretical dynamics Stokes shift, $\Delta\tnu_{\rm theo} = \tnu_{\rm p}(0) - \tnu_{\rm p}(\infty)$, is amenable. We use this additional information to fit the following equation to the experimental peak positions of \C, $\tnu_{\rm p}(t)$
\begin{equation}
\tnu_{\rm p}(t) = (\Delta\tnu_{\rm theo} - \sum_i \Delta\tnu_i)e^{-\frac{t}{\tau_{\rm s}}} + \sum_i \Delta \tnu_i e^{-\frac{t}{\tau_i}} + \tnu_\infty,
\end{equation}
which serves to obtain a fully analytical $C(t)$ for further use in the determination of $D(t)$ for the GSE and of $\eta(t)$ for the GLE. Here we have fixed the shortest lifetime, $\tau_{\rm s}$, to \unit[100]{fs}. Figure~\ref{fig:Ct_C153} shows exemplary experimental data and the corresponding fits.

\subsection{Transient Absorption}

\begin{figure}[!htp]
\centering
	\includegraphics[]{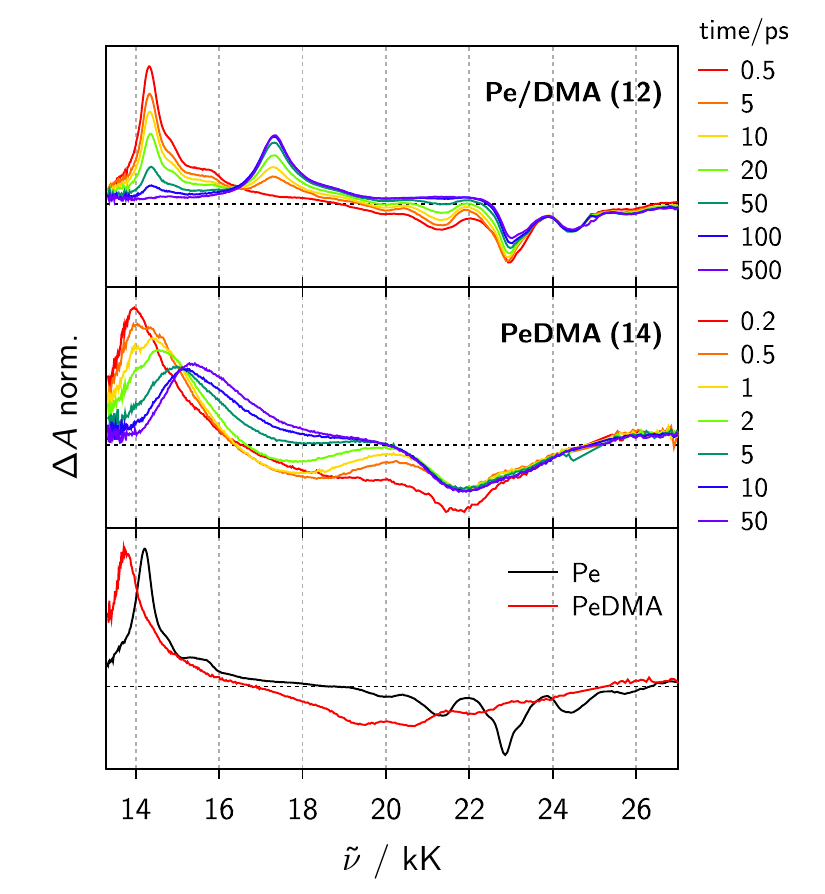}
\caption{Transient absorption (TA) spectra of \Pe\ and \PD\ in {\it c}-hexane after \unit[5]{ps} (lower panel). Selected TA spectra of \PD\ in dimethylsulfoxide (middle panel) and of \Pe\ and \DMA\ (\unitfrac[0.8]{mol}{L}) in acetonitrile (upper panel).}
\label{fig:ta_spectra}
\end{figure}

Figure~\ref{fig:ta_spectra} compares transient absorption spectra of \PD\ and \Pe\ in apolar {\it c}-hexane, as well as in a polar solvent with \Pe\ in the presence of an excess of \DMA. The transient difference spectra of \PD\ and \Pe\ in an apolar solvent show a substantial resemblance to each other with the most noticeable differences being the red-shifted and broadened ground-state bleach (GSB) and stimulated emission, in agreement with our findings in the steady-state spectra. The excited state absorption of \PD\ at approx.\ \unit[14]{kK} is almost identical to the one observed in pure \Pe, though slightly more broadened. In a polar solvent, however, the situation changes drastically. While the freely diffusing \Pe/\DMA\ pair almost perfectly produces precursor-successor kinetics, with an isosbestic point in the TA spectra at approx.\,\unit[16.5]{kK}, the linked \PD-system shows a smooth transition from a spectrum resembling the one in cyclohexane to a broad absorption band peaking at approx.\,\unit[15.5]{kK}. This band is more than \unit[2]{kK} red-shifted with respect to the \Pe-anion band. The absence of an isosbestic point is even more appreciated when the stimulated emission is subtracted from the TA spectra, leaving only excited state absorption features and the GSB (see Fig.~S5).


\section{Modelling \& Simulations}
\subsection{Free Energy Surfaces}

The steady-state solvatochromism, transient absorption data, as well as the fact, that freely diffusing \Pe\ and \DMA\ can easily undergo electron transfer leads us to opt for a three state model for \PD. This model for calculating the free energy surfaces builds on the following three (diabatic) states: a) \Pe\ and \DMA\ in their respective electronic ground state, b) the locally electronically excited \Pe\ and \DMA\ in its ground state and c) the fully charge separated state with the \Pe\ anion and \DMA\ cation, which we will label as $\ket{\rm g}$, $\ket{\rm l}$ and $\ket{\rm c}$.
\begin{align}
\ket{\rm g} & = \ket{\rm PD} \nonumber\\
\ket{\rm l} & = \ket{\rm P^*D} \\
\ket{\rm c} & = \ket{\rm P^{\bullet-}D^{\bullet+}}.\nonumber
\end{align}
The matrix representation of the Hamiltonian, $\bf H$, and the dipole moment matrix, $\mu$, in this diabatic basis are given as:\cite{kang_CP_1990, kattnig_PCCP_2011}
\begin{equation}
\bf{H}(\mu_{\rm s}) = \left(\begin{array}{ccc}F_{\rm g}(\mu_{\rm s}) & 0 & J_{\rm gc} \\0 & F_{\rm l}(\mu_{\rm s})  & J_{\rm lc}  \\J_{\rm cg}  & J_{\rm cl}  & F_{\rm c}(\mu_{\rm s}) \end{array}\right),
\label{eq:H}
\end{equation}
and
\begin{equation}
\mathbf{\mu} = \left(\begin{array}{ccc}\mu_{\rm g}  & \mu_{\rm gl} & \mu_{\rm gc} \\ \mu_{\rm lg} &  \mu_{\rm l}  & 0  \\\mu_{\rm cg}  & 0 &  \mu_{\rm c} \end{array}\right).
\end{equation}
Here, we assume that {\bf H} and $\mathbf{\mu}$ are symmetric matrices and that the the diagonal elements, $F_i$ in eq.\ \eqref{eq:H}, depend exclusively on the solvent polarization, i.e.\ the instantaneous effective dipole moment, $\mu_{\rm s}$, and can be described within a continuum model for solvation following the work by van der Zwan and Hynes\cite{zwan_JPC_1985}
\begin{equation}
F_i(\mu_{\rm s}) = U_i -  \frac{B_i^{\rm tot}}{2}\mu_i^2 + \frac{B_i^{\rm nuc}}{2}(\mu_i - \mu_{\rm s})^2 + \delta_{0i}B_i^{\rm el}D.
\label{eq:F}
\end{equation}

\begin{figure}[htp]
\centering
	\includegraphics[]{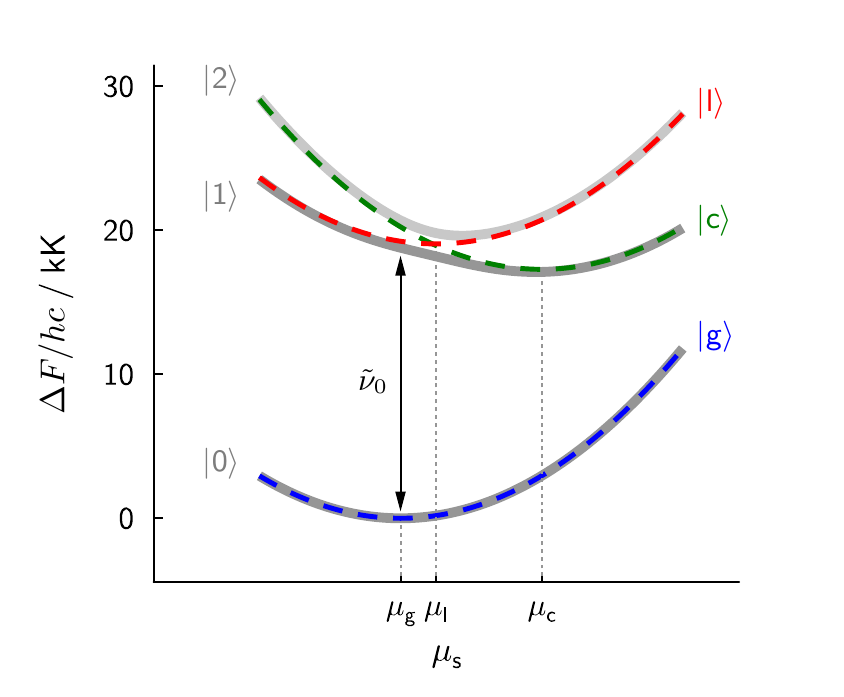}
\caption{Energy scheme depicting the relevant diabatic (dashed blue, red, green) and adiabatic (dark and light grey) states and quantities. Note, that the ordinate has been set to zero by subtracting $(U_{\rm g} - \frac{1}{2}B_{\rm g}^{\rm tot}\mu_{\rm g}^2 + B_{\rm g}^{\rm el}D)/hc$.}
\label{fgr:model}
\end{figure}

This model accounts for the interaction of the permanent point dipole, $\mu_i$, with polarizability, $\alpha_i$, of state $i$ placed in the center of a spherical cavity of radius, $a$, with a continuum dielectric environment, characterized entirely by its static and high frequency dielectric permittivity, i.e.\ $\epsilon$ and $n^2$, respectively. The first term on the right hand side of eq.\ \eqref{eq:F}, $U_i$, accounts for the free energy of state $i$ in the gas-phase, the second one for equilibrium solvation and the third one for the non-equilibrium contribution to the free energy, due to orientational solvent relaxation. The fourth term accounts for the difference in the dispersion interactions between solute and solvent between the excited states and the ground state.\cite{liptay_ZNAPS_1965} In order to keep the number of experimentally inaccessible (or difficult to access) parameters low we have opted for adding the effective last term to the ground state free energy, rather than assigning excited and ground state stabilization energies separately and taking their difference.\footnote{This is of course tantamount to tacitly assuming that the energetic stabilization due to dispersion interactions is the same for both excited states.} In doing so the results from ref.\,\citenum{liptay_ZNAPS_1965} for the absorption and emission transition energies and their solvent dependence are fully recovered.

The solvent response functions, $B_i^{y}$, are given by\cite{liptay_ZNAPS_1965, zwan_JPC_1985, dahl_JPCB_2005}
\begin{subequations}
\label{eq:B}
\begin{align}
B_i^{y} & = \frac{2}{a^3}\cdot\frac{f_x}{1 - 2 (\alpha_i/a^3) f_x} \qquad f_x = \frac{x - 1}{2 x + 1} \\
B_i^{\rm nuc} & = B_i^{\rm tot} - B_i^{\rm el}. \label{eq:Bnuc}\\
(x & = \epsilon \to y = {\rm tot}) \wedge (x = n^2 \to y = {\rm el}) \nonumber
\end{align}
\end{subequations}
$J_{ij}$ in eq.~\eqref{eq:H} denotes the electronic coupling elements between states $i$ and $j$, all of which - in analogy to the treatment in refs.\ \citenum{kang_CP_1990, tominaga_JPC_1991} - are assumed to be equal.\\
Diagonalization of eq.~\eqref{eq:H} at each point along the solvation coordinate $\mu_{\rm s}$ yields three adiabatic states, which we henceforth shall label as $\ket{0}$, $\ket{1}$ and $\ket{2}$. Applying the same basis transformation to the dipole moment matrix, yields the adiabatic permanent dipole moments (diagonal elements) and transition moments (off-diagonal elements).

With the above presented model it is possible to calculate the FES of \PD\ in any solvent of interest, given that key solute (and solvent) properties are known, which are summarized in Table~\ref{tab:pars}. The solvents are entirely parametrized by their static dielectric constant, $\epsilon$, and refractive index, $n$. However, the 16 solute parameters constitute a rather large number. Below we outline how we can reduce this number and determine the remaining parameters either by quantum mechanical calculations or by comparison with experiment.

\subsubsection{Parameters by Assignment}
First we applied some simplifications by assuming that the couplings between the diabatic states involving the charge separated state are all the same, i.e.\ $J_{{\rm c}j} = J_{j{\rm c}} = J$, and that the polarizabilities of all electronic states are identical, i.e.\ $\alpha_i = \alpha$.\footnote{See page 1456 in ref.~\citenum{liptay_ZNAPS_1965} for a discussion on the validity of this assumption.} Secondly, as all observables of interest are related to free energy differences between the involved states, we can set the gas-phase energy of the ground state to zero, i.e.\ $U_{\rm g} = 0$. Finally, the transition connecting states $\ket{g}$ and $\ket{l}$ is essentially nothing else than the first electronic transition of \Pe. Thus, we chose to fix the corresponding diabatic transition dipole moments $\mu_{\rm lg} = \mu_{\rm gl}$ to the value obtained for \Pe. 

\subsubsection{Parameters from Quantum Mechanical Calculations}
It is advisable to rely on quantum mechanical calculations for those parameters in Table~\ref{tab:pars} which are difficult to access experimentally. We thus decided to apply an approach identical to the one outlined in ref.\ \citenum{arzhantsev_JPCA_2006} for determining $\mu_{\rm g}$, $\alpha$ and $a$:
\begin{itemize}
\item The electronic dipole moment in the electronic ground state, $\mu_{\rm g}$, was calculated using Gaussian (gas-phase, B3LYP / 6-31G-(d,p)). 
\item The electronic polarizability of \PD\, $\alpha$, is estimated from an empirical relationship with the van der Waals volume, $V_{\rm vdW}$,\cite{dahl_JPCB_2005}
\begin{equation}
(\alpha/\text{\AA}^3) \cong 0.0268 {(V_{\rm vdW}/\text{\AA}^3)}^{1.35},
\end{equation}
where $V_{\rm vdW}$ was calculated using volume increments.\cite{edward_JCE_1970} 
\item The hydrodynamic cavity radius of \PD, $a$, was determined calculating the permanent electric dipole moment in a series of solvents of different dielectric constant (see Fig.~S6) and applying the following relationship
\begin{equation}
\frac{\mu(\epsilon = 1)}{\mu(\epsilon)} =  1 - 2 \frac{\alpha}{a^3} f_\epsilon,
\end{equation}
where $f_\epsilon$ is defined in eq.\,\eqref{eq:B}. 
\end{itemize}

\subsubsection{Parameters from Solvatochromism}
Six of the seven remaining unknown parameters ($U_{\rm l}$, $U_{\rm c}$, $\mu_{\rm l}$, $\mu_{\rm c}$, $D$, $J$) can be obtained from comparing the experimental steady-state solvatochromism of \PD\ with the one simulated from the FES-model. To obtain the simulations we follow an approach outlined by Maroncelli and co-workers.\cite{horng_JPC_1995} The energy difference, $\tnu_0$, between $\ket{0}$ and $\ket{1}$ at a given position of the solvent coordinate $\mu_{\rm s}$ constitutes the 0-0 transition energy at this specific solvent polarization. In addition, the population distribution in state $i$, $\rho_{i}(\mu_{\rm s})$, and the transition dipole moment of interest, $\mu_{10}(\mu_{\rm s})$, have to be taken into account. Simulated absorption and emission spectra in any given solvent can then be calculated from
\begin{widetext}
\begin{subequations}
\label{eq:obs}
\begin{align}
A(\tnu) & \propto \tnu \int \rho_0(\tnu_0,0)  \mu_{10}^2(\tnu_0) L_{\rm a}(\tnu - \tnu_0) {\rm d}\tnu_0\\
I(\tnu,\infty) & \propto \tnu^3 \int \rho_1(\tnu_0,\infty) \mu_{10}^2(\tnu_0) L_{\rm f}(\tnu - \tnu_0) {\rm d}\tnu_0,
\end{align}
\end{subequations}
\end{widetext}
where $\tnu_0$ denotes the free energy difference between $\ket{0}$ and $\ket{1}$. In eq.~\eqref{eq:obs} we have already undertaken the transformation from the $\mu_{\rm s}$-space to the $\tnu_0$-space making use of the fact that the integral of the transition moment weighted population distribution, $\rho_i(x) \mu_{10}(x)^2$, is a conserved quantity, i.e.\
\begin{equation}
\rho_i(\tnu_0) \mu_{10}^2(\tnu_0) = \rho_i(\mu_{\rm s}) \mu_{10}^2(\mu_{\rm s}) \dfrac{{\rm d}\mu_{\rm s}}{{\rm d}\tnu_0},
\end{equation}
where $\mu_{10}$ denotes the adiabatic transition dipole moment for the $1 \leftarrow 0$ transition. The population distributions for obtaining steady-state spectra are simply given as
\begin{subequations}
\label{eq:rho}
\begin{align}
\rho_0(\mu_{\rm s}) & =  N_0 \exp\left(-F_0(\mu_{\rm s}) / k_{\rm B}T \right) \label{eq:rho0}\\
\rho_1(\mu_{\rm s},\infty) & = N_1 \exp\left(-F_1(\mu_{\rm s}) / k_{\rm B}T \right). \label{eq:rho1}
\end{align}
\end{subequations}
The normalization factors, $N_i$, ensure area-normalized population distributions. It is necessary to note, that for emission this is only true if the equilibrium distribution in the excited state is attained much faster than the actual lifetime of the excited state, which is generally true for conventional organic solvents and solutes with lifetimes in the ns-regime (as is the case for \PD).

The missing solute parameters can now be obtained by nonlinear optimization to reproduce the steady-state absorption and emission spectra in a total of 14 organic solvents and 10 binary solvent mixtures of varying dielectric constant.\footnote{The protic solvents (30-34) were not used for the solvatochromic fitting.} Rather than fitting the entire spectra, we opted for reproducing the first and second moments of absorption and emission, as defined in eq.~\eqref{eq:mom}. The simulated spectra were calculated using eqs.~\eqref{eq:L}, \eqref{eq:obs} and \eqref{eq:rho}.

Finally, the transition dipole moment connecting $\ket{g}$ and $\ket{c}$, $\mu_{\rm cg}$, was adjusted separately until good agreement between the simulated and experimental transition dipole moments for absorption and emission, $\mu_{10}$ and $\mu_{01}$, was achieved. The initial solvatochromic fitting procedure was then repeated with this new parameter, the results of which served as new input for the estimation of $\mu_{\rm cg}$, and so on until convergence was achieved. It is, however, to be noted, that the impact of $\mu_{\rm cg}$ on the spectral moments is rather small.

\begin{figure}[!htp]
\centering
	\includegraphics[]{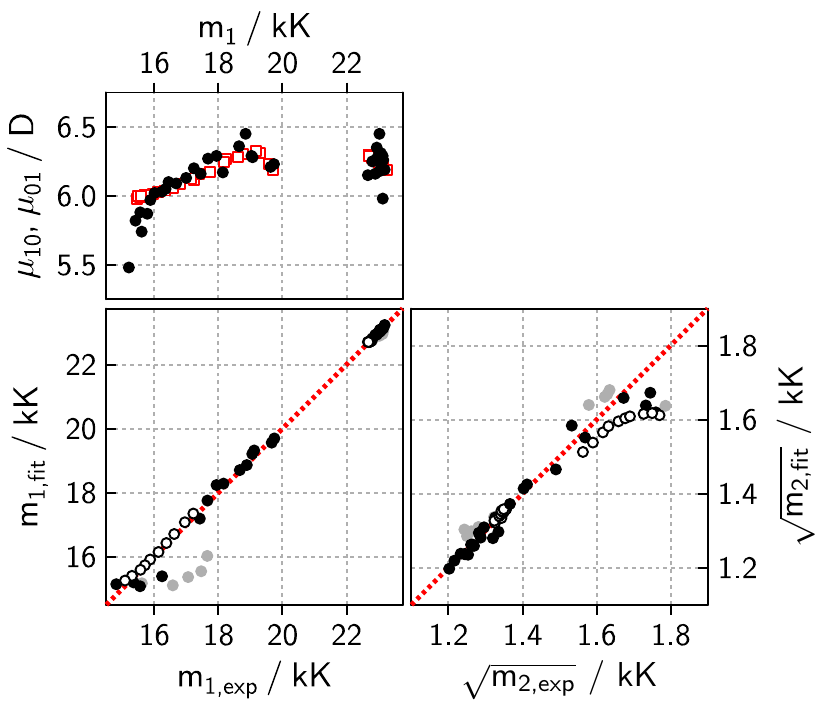}
\caption{Simulated vs.\ experimental first moments of absorption and emission spectra using the parameters given in Table \ref{tab:exps}, upon fitting solvents 1-28. Full black circles ($\bullet$) denote (aprotic) pure solvents (1-14), white circles ($\circ$) are binary solvent mixtures (20-28) and grey circles (\textcolor{lightgray}{$\bullet$}) denote protic solvents (30-34). The upper left panel shows the experimental (black circles) absorption and emission transition dipole moments as a function of the first moment of the corresponding transition. The simulated values are shown as red open squares and are plotted vs.\ the calculated first moments.}
\label{fig:ss_fit}
\end{figure}

Table~\ref{tab:pars} and Figure~\ref{fig:ss_fit} summarize the optimization parameters and compare the calculated and experimental spectral moments and transition dipole moments in all solvents studied (exemplary simulated and experimental spectra are shown in Fig.~S4). We excluded all protic solvents from the analysis for two reasons. First, we do not know whether the solvatochromism of \PD\ in protic solvents can be fully described by using a model based on non-specific interactions. Second, protic solvents tend to exhibit long solvent relaxation times, thus potentially lifting the restriction (eq.~\eqref{eq:rho1}) we had imposed for calculating the emission spectra.

Some of the fitting parameters are worth discussing. Note, that the gas-phase 0-0 transition energy of the locally excited state, $E_{00} = U_{\rm l} - U_{\rm g}$, is only slightly lower in energy by \unit[200]{cm$^{-1}$} than the experimental gas-phase (argon at \unit[20]{K}) value of pure \Pe\ (\unit[23.1]{kK}).\cite{joblin_JCP_1999} Similarly, the permanent dipole moment of the diabatic ''charge-transfer`'' state, $\mu_{\rm c}$, very closely resembles the value that would result from fully displacing a single electron over the center to center distance from \Pe\ and \DMA\ within \PD\ (approximately \unit[6.6]{\AA}). On the downside, we do recognize that the locally excited state diabatic permanent dipole moment, $\mu_{\rm l}$, is not very well defined and the estimated uncertainty allows for a large variation of it. The large coupling, $J$, being more than $6k_{\rm B}T$, is sufficiently large to justify the GLE (and GSE) approach, which relies on adiabatic FESs and that nonadiabatic effects on the excited state dynamics can safely be neglected.

\begin{table}[htp]
\begin{threeparttable}[c]
     \begin{ruledtabular}
     	\caption{\PD-parameters establishing the FES in solvents of arbitrary $\epsilon$ and $n$.}
	\label{tab:pars}
		\begin{tabular}{c...}
			& \multicolumn{3}{c}{state-specific} \\
			 			& \multicolumn{1}{c}{$i = {\rm g}$} & \multicolumn{1}{c}{$i = {\rm l}$} & \multicolumn{1}{c}{$i = {\rm c}$}\\
		\cline{1-4}
		\rule{0pt}{4ex}
		 $\mu_i$ / D		& 3.26 & 5.8 	& 33.2 \\
		$\mu_{i{\rm g}}$ / D 	& 	  & 4.8 	& -2.2 \\
		$U_i/hc$ / kK 	& 0     & 22.9 	& 27.0 \\[1ex]
			& \multicolumn{3}{c}{common}\\
		\cline{2-4}
		\rule{0pt}{4ex}
		$J/hc$ / kK 			& 	  &	1.3	&	 \\
		$\alpha$ / \AA$^3$   	&   	&	71	&	\\
		$a$ / \AA			&   &	7.76	&	\\
		$2D/hca^3$ / kK			&   &	3.9	&	
		\end{tabular}
     \end{ruledtabular}
\end{threeparttable}
\end{table}

The FES can also be rewritten in normalized reaction coordinates, $z$, to facilitate the simulations with the GLE. The transformation is straightforward
\begin{equation}
z = \frac{\mu_{\rm s} - \mu_{\rm l}}{\Delta \mu}
\end{equation}
where $\Delta \mu = \mu_{\rm c} - \mu_{\rm l}$. This yields the final equations for the FES in this new coordinate as
\begin{equation}
F_i(z) = U_i - \frac{B^{\rm tot}}{2}\left(z_{i,{\rm eq}}\Delta\mu + \mu_{\rm l} \right)^2 + \frac{B^{\rm nuc}}{2} \left(\mu_i - \mu_{\rm l} - z\Delta\mu \right)^2,
\end{equation}
where $z_{i,{\rm eq}}$ denotes the equilibrium position of state $i$.

\subsection{Dynamic Quantities}

In both models the GLE and the GSE, the dynamics are represented by either the friction kernel, $\eta(t)$, or the time dependent diffusion coefficient, $D(t)$.\footnote{To the best of our knowledge, even in the case of the harmonic oscillator, the relationship between these two quantities is only defined via Laplace transforms.} Assuming, that to a large extent both time dependent functions are independent of the FES over which the solute evolves, i.e.\ independent of the nature of the solute and exclusively a function of the solvent, both quantities can be obtained by measuring the solvation dynamics in a free energy potential of known - and if possible simple - form. Two-level systems, such as charge-transfer dyes like coumarins, fulfill this condition and - in medium to polar solvents - exhibit a reasonably quadratic free energy surface in the ground and excited state.\cite{horng_JPC_1995} Due to this and the reasonably large dynamic Stokes shifts, coumarins have been exhaustively used in the past to reliably report solvation dynamics.\cite{horng_JPC_1995, gustavsson_JPCA_1998, sajadi_JPCB_2013, kumpulainen_PCCP_2017} Here we have used the well-studied \C, the spectral response function, $C(t)$, of which, is supposed to provide a reasonable experimental equivalent of the solvation energy response and thus the basis for evaluating $D(t)$ and $\eta(t)$.\cite{maroncelli_JML_1993}

\subsubsection{Time-dependent Diffusion Coefficient}

Following van der Zwan and Hynes, the time-dependent diffusion coefficient in the GSE can be obtained from the solvent relaxation dynamics, $C(t)$ in a harmonic FES as follows\cite{zwan_JPC_1985}
\begin{equation}
D(t) = -\dfrac{k_{\rm B}T}{B^{\rm nuc}}\dfrac{\dot{C}(t)}{C(t)},
\label{eq:Dt}
\end{equation}
where $C(t)$ (eq.\,\eqref{eq:Ct}) and $B^{\rm nuc}$ (eq.\,\eqref{eq:Bnuc}) have been defined before.

\subsubsection{Time-dependent Friction}

As mentioned above, the friction kernel, $\eta(t)$, in the GLE is related to the solvent dielectric relaxation function of the solvent, which can be measured experimentally by observing the time dependent peak shift of a polarity probe like \C. According to Hynes,\cite{zwan_JPC_1985} the solvent correlation function can be related to the properties of the stochastic variable for a pure harmonic potential well:
\begin{equation}
C(t) \approx \Delta(t) = \dfrac{\braket{z(0)z(t)}}{\braket{z^2}}.
\end{equation}
Upon Laplace transforming the Langevin equation in the harmonic potential after multiplying by $z(0)$ and ensemble averaging, one can obtain the friction kernel evaluating the inverse Laplace transform of 
\begin{equation}
\tilde{\eta}(s) = \dfrac{(s^2 + \omega_{\rm L}^2)\tilde{\Delta}(s)-s}{1 - s\tilde{\Delta}(s)},
\label{eq:eta_s1}
\end{equation}
which for a multiexponential function of $C(t)$ takes the form of:
\begin{equation}
\tilde{\eta}(s) = \dfrac{(s^2 + \omega_{\rm L}^2) \sum_i \dfrac{a_i}{s + 1/\tau_i} - s}{1 - s \sum_i \dfrac{a_i}{s + 1/\tau_i} },
\label{eq:eta_s2}
\end{equation}
where the $a_i = \Delta\tnu_i/\sum_i \Delta \tnu_i$ and $\tau_i$ are the result from the fits to the solvation dynamics of \C\ (see Table~SII). The friction kernel in the time domain is evaluated by numerically inverting the above expression using the Gaver-Stehfest algorithm.\cite{stehfest_CA_1970} This result is then fit by the expression:
\begin{equation}
\eta(t) = \omega_{\rm L}^2\gamma\delta(t) + \omega_{\rm L}^2 \sum_i k_i \exp (-\lambda_i t).
\label{eq:eta_t}
\end{equation}
This function is suitable for deriving the set of equivalent ODEs to the GLE (see eqs.~\eqref{eq:gle_split}). Further details are provided in the following section and the supporting information. Equations \eqref{eq:eta_s1} and \eqref{eq:eta_s2} are to be dealt with caution, as it is assumed that the derivative of the solvent correlation function at time zero is zero (otherwise it appears summing in the numerator) in agreement with the fact that the system is pumped  to the excited state from an equilibrium distribution. However, for \C, we make use of a $C(t)$ with a multiexponential decaying function which does not fulfill this condition. On the other hand, this condition is fulfilled if that function is a series of exponential terms plus a series of exponential terms multiplied by sines and cosines. Thus, for the sake of simplicity of fitting we assume that the oscillatory part of the correlation function is small enough to be neglected within the time scale under study, such that we obtain only a sum of exponentials.This is supported by the fact that no oscillations have been observed in our experiments.

There are two parameters in eq.\ \eqref{eq:eta_t} which are not fitted, namely $\omega_{\rm L}$ and $\gamma$. $\omega_{\rm L}$ is the frequency associated to the mass $m_{\rm L}$ in eq.~\eqref{eq:gle}. For chemical reactions, the particle which is moving over the potential energy surface is not a mass, properly speaking. Rather it is related to the change of polarization of the environment, which enables the reaction. In other words, for an electron transfer reaction, the system is suffering energy fluctuations due to the thermal noise that provokes constant changes in the electric field of the medium. In the SI we point out, how $m_{\rm L}$ and $\omega_{\rm L}$ can be obtained.

On the other hand, the $\gamma$-parameter is the equivalent of the Debye linear solvent relaxation time. However, instead of taking this parameter from tables, as we are not dealing with Debye solvents we adjust it in such a way that the obtained friction kernel reproduces the experimental solvent correlation function of \C\ (using the parameters given in Table~S1 for the FES of \C). In doing so, a self-consistent result is obtained. Finally, the so obtained values for $\omega_{\rm L}$, $\gamma$, the $k_i$s and $\lambda_i$s are subsequently used for \PD\ in evaluating the GLE (see Table~SIII).\\

\subsection{(Stochastic) Differential Equations}

\subsubsection{General Smoluchowski Equation}

For a harmonic potential the GSE (eq.\,\eqref{eq:GSE}) can be deduced from the GLE (eq.\,\eqref{eq:gle}) for the temporal evolution of the probability density, $\rho(z,t)$, along the free energy surface $F(z)$ with the time-dependent diffusion coefficient, $D(t)$ defined in eq.\,\eqref{eq:Dt} and the initial condition given by the equilibrium population distribution in the ground state, eq.\,\eqref{eq:rho0}. As mentioned before, and following the example of references \citenum{tominaga_JPC_1991, tominaga_JPC_1991a, meer_JCP_2000, heisler_JPCB_2009, kondo_JPCB_2009} we take the liberty of using the GSE with a potential different from a perfectly harmonic one. Solving the GSE yields the time-dependent excited state population distribution, $\rho_1(z,t)$, which can be used for further analysis. The GSE has been solved using the {\tt pdepe} function in Matlab.\cite{matlab}\\

\subsubsection{Generalized Langevin Equation}
Below, we briefly present the procedure that allows for numerical integration of eq.~\eqref{eq:gle}.\cite{fonseca_CPL_1989, gudowska-nowak_APPB_1994, goychuk_2012, cordoba_JR_2012, baczewski_JCP_2013} The GLE, which describes a process that is non-Markovian due to the presence of memory, can be split into a set of stochastic differential equations, where the memory kernel is not explicitly present (see SI for derivation).\cite{fonseca_CPL_1989, gudowska-nowak_APPB_1994, goychuk_2012, cordoba_JR_2012, baczewski_JCP_2013} Each of the equations  describes the Markovian evolution of an auxiliary variable, driven by independent Gaussian white noise:
\begin{widetext}
\begin{align} \label{eq:gle_split}
\dfrac{{\rm d}z (t)}{{\rm d}t} & = v(t) \nonumber \\ 
\dfrac{{\rm d}v (t)}{{\rm d}t} & = -\dfrac{1}{m_{\rm L}} \dfrac{\partial F(z)}{\partial z} - \omega_{\rm L}^2\gamma v(t) + \sum_i w_i(t) + \omega_{\rm L}^2 \sqrt{\dfrac{\gamma k_{\rm B}T}{\lambda_{\rm s}}} \xi_0(t)\\ 
\dfrac{{\rm d}w_i (t)}{{\rm d}t} & = -\lambda_i w_i(t) - \omega_{\rm L}^2k_i v(t) + \omega_{\rm L}^2 \sqrt{\dfrac{\lambda_i k_i k_{\rm B}T}{\lambda_{\rm s}}}\xi_i(t) \nonumber 
\end{align}
\end{widetext}
In Eqs. \ref{eq:gle_split}, the variables $\xi_j(t)$, $j=0\dots N$, denote independent Gaussian noises with correlations:
\begin{subequations}
\begin{align}
 \braket{\xi_j(t)} &= 0, \\ 
 \braket{\xi_j(t) \xi_j(t+\tau)} &= \delta(\tau), 
\end{align}
\end{subequations}
The number of auxiliary variables, $w_i$, corresponds to the number of exponential components in the friction kernel. The variables $w_i$ have the dimension of ``acceleration'' of the $z$ variable.

The initial conditions for this set of equations can be obtained assuming that the system starts from an equilibrium Gaussian distribution, which is a projection to the excited state of that in the ground state. Under such conditions the velocity is zero. The second derivative of the stochastic variable, the acceleration, is equal to the derivative of the potential divided by the inertial mass of the solvent.  Therefore, the sum of the initial values for the other quantities, $w_i$, is also zero.

As there is no reason to believe that in the case of several of these quantities their initial values compensate, they all must be zero at time zero:
\begin{align}
v(0) & = 0\\
w_i(0) & = 0
\end{align}
The initial distribution of the stochastic variable comes from the distribution of the system at rest in the ground state
\begin{equation}
\rho_1(z,t=0) = \rho_0(z,\text{eq}) = N \exp \left(-F_0(z)/k_{\rm B}T \right),
\end{equation}
where $N$ ensures area-normalized population distributions.

Generalized Langevin equations can be solved analytically only in some special cases.\cite{goychuk_2012} Therefore, in order to integrate Eqs.\,\eqref{eq:gle_split} for our system, with the initial  conditions given above, we use the numerical Euler-Maruyama scheme (see SI for explanation of the algorithm).\cite{mannella_IJMPC_2002} For the sake of consistency we have also used the very same approach for \C\ despite having an (almost perfectly) harmonic potential.

\section{Comparison of experiments with simulations}

As explained in the experimental part, the first two moments of the experimental time resolved spectra were extracted. From the simulations we obtain trajectories starting at different points in the LE region of the FES. These trajectories are weighted taking into account the initial distribution in the S$_1$ after excitation, and the histograms which are equivalent to the population evolution $\rho(\tnu_0,t)$ are calculated. Finally, the time-resolved emission spectra are simulated according to
\begin{equation}
I(\tnu,t) \propto \tnu^3 \int \rho_1(\tnu_0,t) \mu_{10}^2(\tnu_0) L_{\rm f}(\tnu - \tnu_0) {\rm d}\tnu_0.
\end{equation}
The lineshape function, $L_{\rm f}$, and the frequency-dependent transition-dipole moment, $\mu_{10}$, are the same as described before and are obtained from the calculation of the FES. The so simulated time resolved spectra are analyzed just as the experimental spectra, i.e.\ the time-dependencies of the first and second spectral moments are calculated. A flow chart summarizing the whole procedure of analysis and comparison of the experiments with the simulation results for the GLE (or GSE) can be found in Fig.~S1. As pointed out before, the solvatochromic analysis leading to the FES is not unambigious, as some parameters are rather ill defined. In order to test how these variations affect our simulations we tested different FES-parameter sets (with comparable goodness of fit in the solvatochromic fitting). Figure~S10 clearly shows, that the differences are negligible.

A comparison of the experimental and simulated (GLE and GSE) results is given in Figs.\,\ref{fig:Ct_pure}-\ref{fig:Ct_visco}. In Fig.~S7 we compare the times required for the dynamics Stokes shift to reach $1/e = 0.368$, $\tau_{1/e}$, for \PD\ and \C.\footnote{For \PD\ we have used the shift of the first moment, $m_1$, while for \C\ we used the peak maximum of the log-normal function, $\tnu_{\rm p}$.} These times are to be taken as an orientation only and not as a quantity with physical significance characterizing the dynamics. The first observation is that in almost all the cases, in the mixtures too, the dynamics are significantly slower for \PD\ than for \C.\footnote{Only at the highest content of glycerol there is a coincidence in the $C(t)$ (see Figs.~S7 and S8) a case in which the dynamics of \C\ are affected by the H-bonding (see section V.B.2 and reference~\citenum{sajadi_PCCP_2011})} This is a direct consequence of the difference in the free energy surface for these two molecules. This is noteworthy, as very often in the literature the solvent dynamic control of a chemical reaction is thought to be active only if the mean or one of the solvent relaxation times is directly reflected in the kinetics of the reaction, which is not necessarily the case.\cite{banerji_JPCA_2008} We have also extracted these times from the GSE and GLE calculations and compared them to the \PD\ times (see Fig.~S7). In the case of the GSE the obtained times deviate more from the experimental values than in the case of the GLE, and increasingly with the magnitude of the relaxation time. In the case of the GLE the largest deviation is observed for the glycerol-rich mixtures. All these deviations are the opposite compared with the \C-times: the $1/e$-times from the GLE and GSE simulations - if off from the experimental values - are larger than the measured ones. These times are to be taken with care as in any case the dynamics are more complex than simple exponential functions. In the coming sections a more detailed comparison is presented.

In addition to the GSE and GLE analysis we have performed a test calculation with the Smoluchowski equation keeping the diffusion coefficient at a constant value and equal to the asymptotic one, reached at long times for two cases: acetonitrile (12) and dimethylsulfoxide (14). The corresponding figure can be found in the SI (Fig.~S9). This is the case of pure overdamping with a memory-less friction. These results allow for two observations: First, for a ``fast'' solvent like acetonitrile the memory is lost much earlier than for dimethylsulfoxide. Second, the overdamped description (Smoluchowski equation) - at least with only a single relaxation mode of the solvent - does not suffice to explain the observations.

\subsection{Pure Solvents}

\begin{figure*}
\centering
	\includegraphics[]{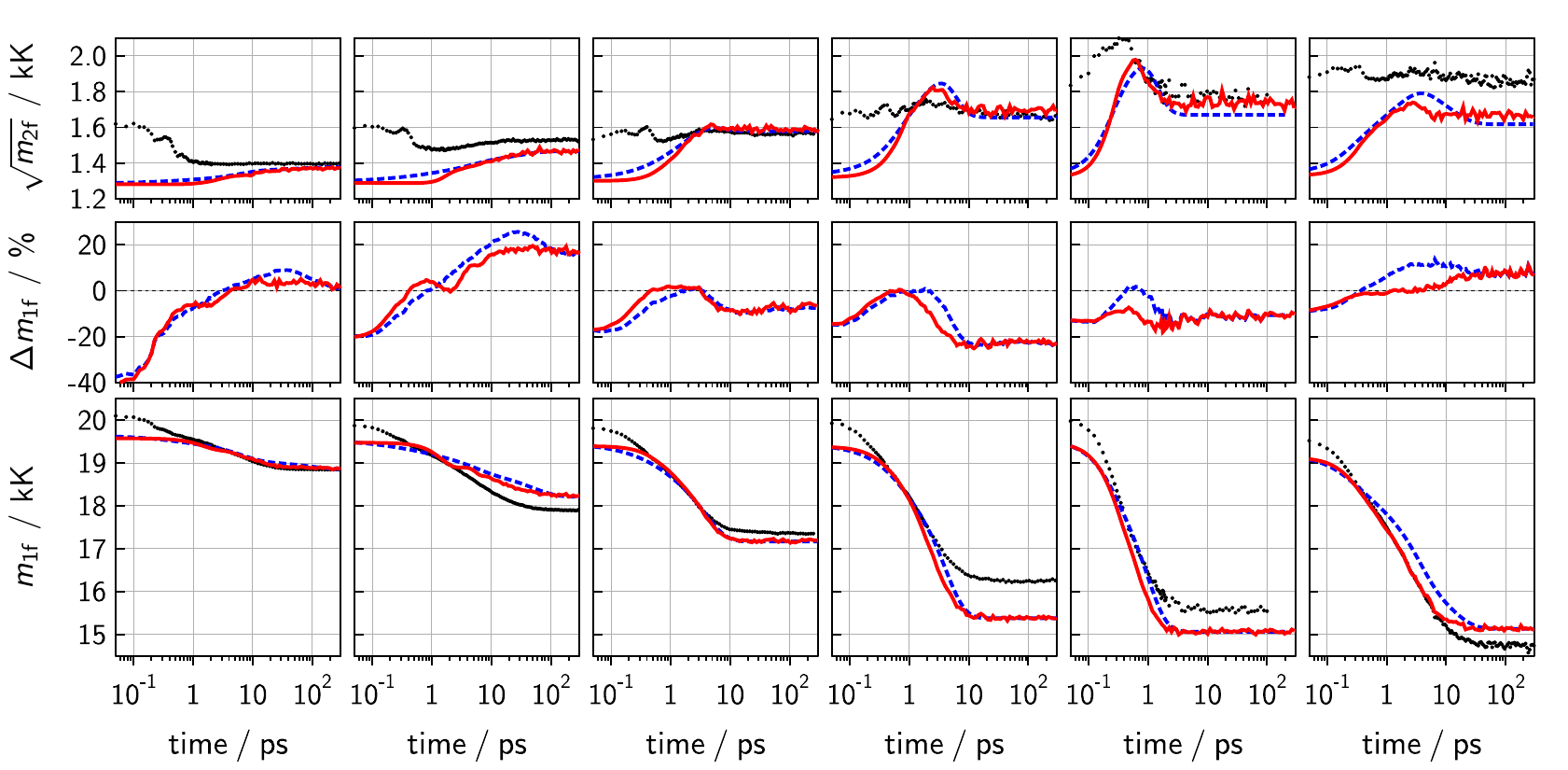}
\caption{Comparison of the experimental (black dots) and simulated (GSE - blue dashed line; GLE - red line) $m_{\rm 1f}$ (lowest row) and $\sqrt{m_{\rm 2f}}$ (upper row) in a set of pure solvents. In addition the relative difference of the simulated and the experimental $m_{\rm 1f}$, given by $\Delta m_{\rm 1f} = (m_{\rm 1f}^{\rm sim} - m_{\rm 1f}^{\rm exp}) / (m_{\rm 1f}^{\rm exp}(t=0) - m_{\rm 1f}^{\rm exp}(t=\infty))$, is shown in the middle row. Solvents used are from left to right: 6, 8, 10, 11, 12 and 14.}
\label{fig:Ct_pure}
\end{figure*}

Fig.~\ref{fig:Ct_pure} represents the mean frequency (first moment) and the standard deviation (square root of the second moment) of the time resolved emission of \PD\ in 6 pure solvents. For the mean frequency we observe a deviation between experiment and simulation in all cases at very short times, typically below \unit[1]{ps}. This deviation is explicitly shown in the middle panel as relative deviation between simulation and experiment, normalized to the full dynamic Stokes shift. The same is true for the standard deviation, $\sqrt{m_{\rm 2f}}$. This is a consequence of the excess vibrational energy after excitation to the $v'=2$ at \unit[400]{nm}. The time-dependence of $\sqrt{m_{\rm 2f}}$ is especially revealing in this respect: due to the displacement between the minima of the S$_0$ and S$_1$ this moment is expected to increase with time, passing through a maximum if the dynamics are fast enough to enlarge the population distribution in the S$_1$ (see for example acetonitrile). The simulations clearly follow this behavior, but not exactly the experiments. In the experiments, the standard deviation tells us that at short times the band is broader than expected. This is a consequence of the emission from vibrationally excited states. This broadening disappears on a time-scale similar to that characterizing the deviations between simulations and experiments of the mean frequency. Beware, that the initial discrepancy between the experimental and simulated standard deviations is always in the range of \unit[0.3-0.5]{kK}. A quite clarifying additional experiment in this respect is the relaxation of the emission in pure {\it c}-hexane in which no charge separation occurs at all (see Fig.\,S11): the dynamics of both moments are almost identical. The total relaxation of the first moment is in the range of \unit[0.4-0.5]{kK}, which is in agreement with the short-time differences observed between simulations and experiments in the other solvents. In addition, the standard deviation of the spectra also changes by about \unit[0.3]{kK} as observed for the other solvents for the very same reason.

In all solvents both the GLE and the GSE reproduce the time scales of the decay of the mean frequency reasonably well.  Only in the case of dimethylsulfoxide the GLE is clearly closer to the experiments than the GSE, which itself is slightly slower than the measured data. In several cases larger discrepancies between the simulations and the experiments are observed at long times. These deviations are a direct result of the deviations from modeling the steady-state solvatochromism (compare with the differences in Fig.\,\ref{fig:ss_fit}). In other words, these deviations cannot be ascribed to imperfections of modeling the dynamics but to deviations from the fits to the stationary spectroscopic data. Note that such deviations would most probably have passed unnoticed if the correlation function (eq.\,\eqref{eq:Ct}) would have been used instead in the data presentation.

\subsection{Solvent mixtures}
\subsubsection{Isoviscous binary mixtures}
\begin{figure*}
\centering
	\includegraphics[]{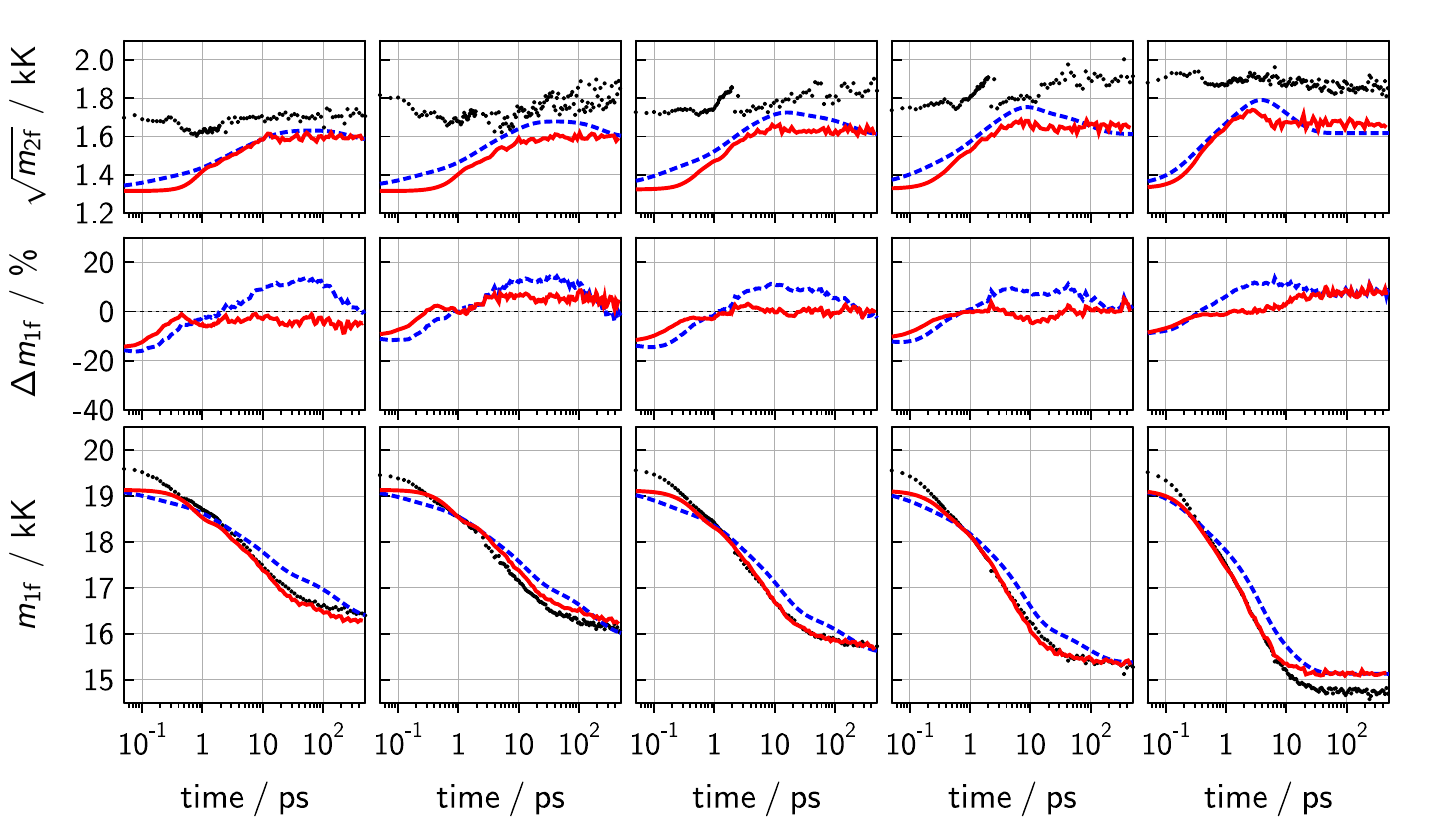}
\caption{Comparison of the experimental (black dots) and simulated (GSE - blue dashed line; GLE - red line) $m_{\rm 1f}$ (lowest row) and $\sqrt{m_{\rm 2f}}$ (upper row) in a set of isoviscous binary mixtures of BzAc/DMSO with changing dielectric constant. The dielectric constants of the binary mixtures are from left to right: 12, 16, 23, 30 and 48}
\label{fig:Ct_elec}
\end{figure*}

For the dynamics in the binary mixtures of benzyl acetate / dimethylsulfoxide a similar trend as in the case of pure solvents is reproduced for both observables (cf.\ Fig.\,\ref{fig:Ct_elec}). Again, the GSE is slightly slower and worse than the GLE, when it comes to reproducing the data for cases where the relaxation is slower. In general it seems that the GSE adds a long component, which is not observed experimentally. We can summarize these findings by saying that the applicability of the GLE seems to be independent of the dielectric properties of the medium and works equally well for solvents of relatively low and high polarity.

\subsubsection{Isoelectric binary mixtures}

\begin{figure*}
\centering
	\includegraphics[]{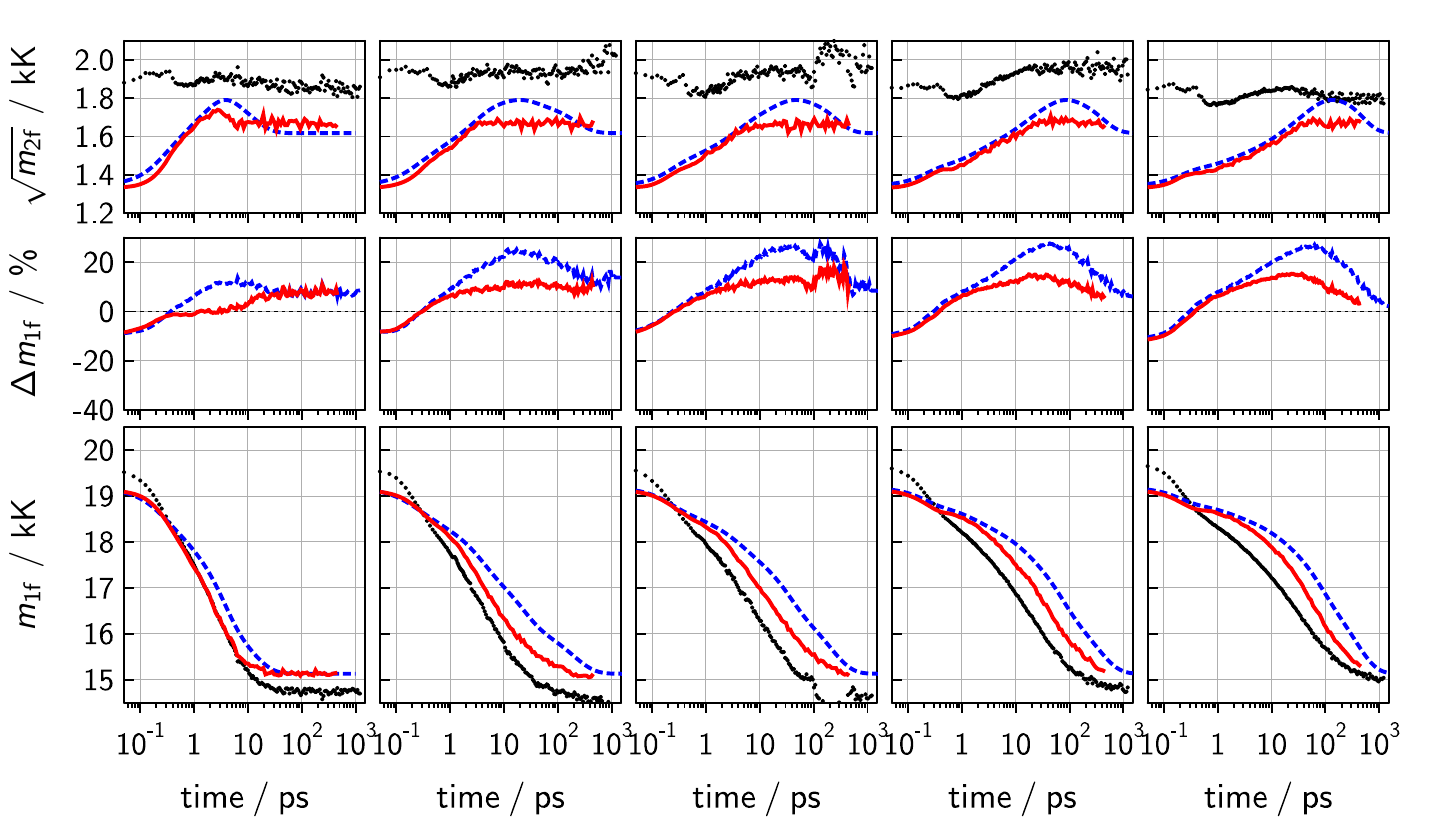}
\caption{Comparison of the experimental (black dots) and simulated (GSE - blue dashed line; GLE - red line) $m_{\rm 1f}$ (lowest row) and $\sqrt{m_{\rm 2f}}$ (upper row) in a set of isoelectric binary mixtures of DMSO/GLY with changing viscosity. The solvent viscosities of the binary mixtures are from left to right: 2, 5, 10, 20 and \unit[40]{cP}.}
\label{fig:Ct_visco}
\end{figure*}

Larger differences between the GSE and the GLE can be observed, when the viscosity is increased while the dielectric properties are kept almost unaltered (cf.\ Fig.\,\ref{fig:Ct_visco}), as is the case in the binary mixtures of dimethylsulfoxide and glycerol. As soon as the viscosity is increased to \unit[5]{cP} the GSE shows significantly slower dynamics. The GLE, on the other hand, is still able to reproduce the experimental results quite well, except for the largest of the viscosities. However, we have the impression that this deviation is again not due to a failing model for the dynamics but a consequence of extracting the friction kernel from another molecule, \C. Upon increasing the viscosity by adding glycerol, the hydrogen bonding ability of the binary solvent mixture changes. More precisely, the $\alpha$-parameter of the Kamlet-Taft (KT) scale increases steadily starting from \unit[20-30]{cP}.\footnote{See Fig. 4 in ref.~\citenum{angulo_PCCP_2016}} \C\ on the other hand has a corresponding $a^*$-coefficient of \unit[-0.86]{kK} in the excited state,\cite{molotsky_JPCA_2003} while, although unknown, for \PD\ we expect it to be lower: for example this value for another aromatic amine is \unit[-0.34]{kK},\cite{rosspeintner_JPPA_2006} and in \PD\ the charge is transferred to the \Pe\ moiety, thus further decreasing this value. Therefore, a correlation between the difference of the KT $\alpha$ coefficient and the deviation of the simulations with respect to the experiments seems reasonable. This is in line with the conclusions obtained by Ernsting for 4-aminophthalimide in methanol.\cite{sajadi_PCCP_2011} There the authors had attributed the observed differences in the solvent relaxation times to concomitant changes in the rotational relaxation times as a result of specific solute-solvent interactions. It is possible, nevertheless, that whenever the dielectric relaxation is very fast the influence of slower phenomena, like hydrogen-bonding, is not affecting the dynamics. In the present case the dielectric relaxation is much slower, opening the possibility for hydrogen-bonding to differently affect the dynamics of \C\ and \PD, thus invalidating the use of the friction kernel obtained with the former to explain the results of the latter.

\section{Conclusions}
In this paper we have shown how, with very elementary models for both the thermodynamics and the non-equilibrium dynamics of solvation, it is possible to describe all experimental observables for a model compound in which a charge transfer reaction occurs in the excited singlet state. The FES are obtained from fitting a three-state model based on Liptay's continuum theory for equilibrium solvation to the steady-state absorption and emission spectra. Moreover, there are almost no free parameters in the simulation of the dynamics of the model compound, \PD, as the relevant solvent parameters are related to those obtained from measurements of \C, a well established reference in this kind of studies. To the best of our knowledge so far no such comprehensive study combining ultrafast broadband spectroscopy and the GLE has been performed. The deviations of the GLE with respect to the experiments are ascribed to vibrational relaxation at short times, to short-comings of the solvatochromic model at long times, and to differences in the hydrogen-bonding sensitivity  for \C\ and \PD\ in glycerol-rich binary mixtures, but not to the intrinsic capabilities of the GLE in describing the dynamics. In general both used models (GLE and GSE) work relatively well as long as the energy dissipation is fast enough, but the GSE fails in those cases where the relaxation becomes slower. A plausible explanation for this difference when the dynamics are slower is that in such cases the population in the S$_1$-state has more time to explore all the details of the shape of the FES. In other words, the differences between the harmonic potential and the one used for \PD\ become more relevant, and the assumption for the validity of the GSE is removed.

Two possible improvements can be foreseen: an improved stationary solvation model to encompass the molecular details of the \cite{small_JACS_2003, matyushov_JPCA_2017} and/or the addition of another reaction coordinate to the dynamics to account for vibrational relaxation and other internal modes of freedom. So far we have only considered the polarization of the medium. In this sense, both the groups of Hynes\cite{malhado_JPCA_2011} and especially Hammes-Schiffer\cite{schwerdtfeger_JCP_2014}, and Ivanov,\cite{yudanov_RJPCB_2013} the latter in the context of the Smoluchowski approach, but expliciting vibrational modes, have performed substantial theoretical advances. Further work in this direction is, without any doubt, necessary when comparing full sets of experimental data as presented here.

The conventional approach for extracting the solvent relaxation times from experimental $C(t)$s consists in fitting them with a multiexponential decay function. However, it is unclear whether the functional shape of an intrinsically noisy set of data such as the $C(t)$ necessarily is a sum of exponentials, and if - even in such a case - each of the exponents corresponds to a single relaxation time of the solvent or is a manifestation of memory effects of this relaxation. With the present approach we have shown that it is possible to explain the dynamics of relaxation of the fluorescence by using a single relaxation mode for the solvent with memory, instead of several times. It's nevertheless unquestionable that further comparison with dielectric spectroscopy data would be needed to establish the microscopic physical origin in each of the solvents of the energy dispersion or the solvent response to perturbations.


\begin{acknowledgments}
AR thanks Niko Ernsting for help with the FLUPS, Eric Vauthey and Bernhard Lang for continued support and discussions, and acknowledges financial support from the SNF (IZK0Z2\_170389). GA and PP would like to acknowledge financial support from the Narodowe Centrum Nauki (Poland) through the grant SONATA bis number 2013/10/E/ST4/00534. GA cherishes the helpful discussions with Tomasz Karda\'s from IPC/PAS.
\end{acknowledgments}

%

\pagebreak
\widetext
\begin{center}
\textbf{\large Supplemental Material: How good is the generalized Langevin equation to describe the dynamics of photo-induced electron transfer in fluid solution?}
\end{center}
\setcounter{equation}{0}
\setcounter{figure}{0}
\setcounter{table}{0}
\setcounter{page}{1}
\makeatletter
\renewcommand{\theequation}{S\arabic{equation}}
\renewcommand{\thefigure}{S\arabic{figure}}
\renewcommand{\thetable}{S\arabic{table}}
\renewcommand{\bibnumfmt}[1]{[S#1]}
\renewcommand{\citenumfont}[1]{S#1}

\setcounter{section}{0}
\renewcommand{\thesection}{S\arabic{section}}
\renewcommand{\thesubsection}{\thesection.\Alph{subsection}}


\begin{figure}[htp!]
	\includegraphics[scale=0.9]{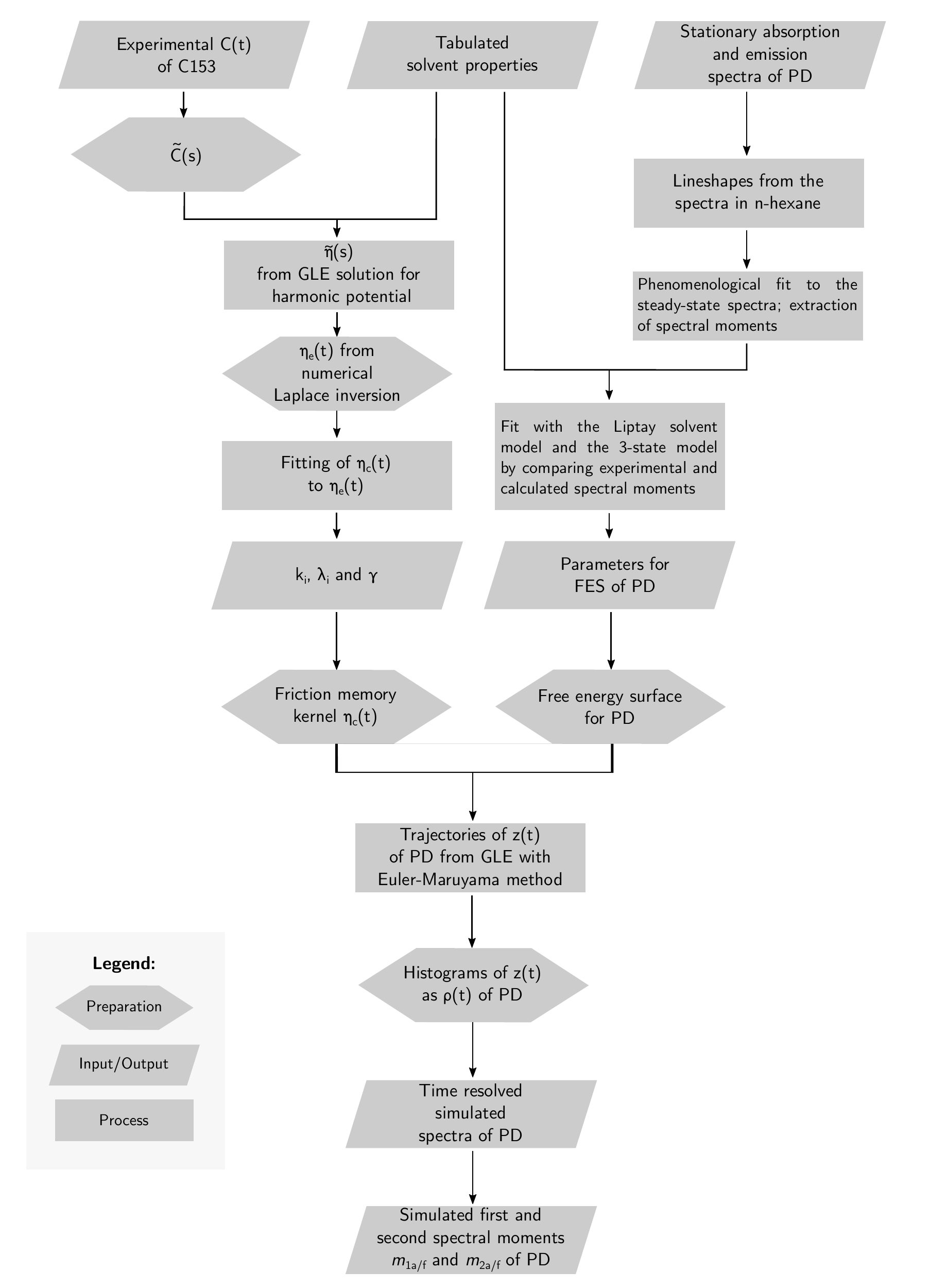}
\caption{Flowchart of the essential experimental and calculational steps used in the manuscript.}
\label{fig:}
\end{figure}

\tocless\section{Correlation function}

We begin with the relationship between $\tilde{\eta}(s)$ and $\tilde{\Delta}(s)$ which follows from Eq. (\ref{eq:gleSI})
\begin{equation}
\tilde{\eta}(s) = \dfrac{(s^2 + \omega_{\rm L}^2)\tilde{\Delta}(s)-s - c}{1 - s\tilde{\Delta}(s)},
\label{eq:eta_s1 app}
\end{equation}
where $c = \dot{\Delta}(0)$. From Eq. (\ref{eq:eta_s1 app}) we obtain
\begin{equation}
\tilde{\Delta}(s) = \dfrac{\tilde{\eta}(s) + s + c}{s^2 + \omega_{\rm L}^2 + s \tilde{\eta}(s)},
\label{eq:delta_s1 app}
\end{equation}
For the friction kernel of the form     
\begin{equation}
\eta(t) = \omega_{\rm L}^2  \left(\frac{t}{\gamma}\right) + \omega_{\rm L}^2 \sum_{i=1}^{N} k_i \exp (-\lambda_i t)
\label{eq:eta_t}
\end{equation}
we have (note that $\gamma\delta(t) = \delta(t/\gamma)$, but the second form is dimensionally correct)
\begin{eqnarray}
\tilde{\Delta}(s) &=& \dfrac{\omega_{\rm L}^2\gamma + \omega_{\rm L}^2 \sum_{i=1}^{N} \dfrac{k_i}{s + \lambda_i} + s + c}{s^2 + \omega_{\rm L}^2 + s \left(\omega_{\rm L}^2\gamma + \omega_{\rm L}^2 \sum_{i=1}^{N} \dfrac{k_i}{s + \lambda_i} \right)}= \nonumber \\ =\frac{\mathcal{N}(s)}{\mathcal{D}(s)} &=& \dfrac{(\omega_{\rm L}^2\gamma + s + c)\mathcal{P}(s) + \omega_{\rm L}^2 \sum_{i=1}^{N} k_i \mathcal{Q}_i(s)}{\left(s^2 + \omega_{\rm L}^2 + s \omega_{\rm L}^2\gamma \right)\mathcal{P}(s)  + s \omega_{\rm L}^2 \sum_{i=1}^{N} k_i \mathcal{Q}_i(s)} \nonumber \\
\label{eq:delta_s1_app_P_and_Q}
\end{eqnarray}
where 
\begin{equation}
\mathcal{P}(s)=\prod_{i=1}^{N} (s + \lambda_i), ~~ \mathcal{Q}_i(s)=\prod_{j=1, j\neq i}^{N} (s + \lambda_j).
\label{eq:definition_of_P_and_Q}
\end{equation}
Numerator $\mathcal{N}(s)$ of the expression in the second line of Eq. (\ref{eq:delta_s1_app_P_and_Q}) is polynomial of degree $N+1$, whereas denominator $\mathcal{D}(s)$ is a polynomial of degree $N+2$.
Therefore, if $\mathcal{N}(s)$ and $\mathcal{D}(s)$ have no common zeros, $\mathcal{D}(s)$ has exactly $N+2$ roots. If all roots are single and real, we obtain $\Delta(t)$ as a sum of purely exponential terms. However, this is not the most general case, as some roots may have nonzero imaginary part. Assume that denominator has $N_O$ pairs of conjugate complex roots, and $N_D$ single, real roots, $2N_O + N_D=N+2$. 
We assume that all roots are single. Moreover, if $\omega_{\rm L}^2 \prod_{i=1}^N \lambda_i \neq 0$, then $s=0$ is not a root of $\mathcal{D}(s)$.

In such case, (\ref{eq:delta_s1_app_P_and_Q}) can be rewritten as
\begin{equation}
\tilde{\Delta}(s)= \sum_{i=1}^{N_D} \frac{A_i}{s+\tau_{i}^{-1}}  + \sum_{i=1}^{N_O} \frac{b_i s + c_i}{(s+\kappa_i)^2 + \Omega^2_i}
\label{eq:Delta_s_partial_fractions}
\end{equation}
Inverting Eq. (\ref{eq:Delta_s_partial_fractions}) we obtain
%
%
\begin{equation}
\Delta(t)= \sum_{i=1}^{N_D} A_i e^{-t/\tau_i} + \sum_{i=1}^{N_O} e^{-\kappa_i t}\left[B_i \cos(\Omega_i t) + C_i \sin(\Omega_i t) \right], 
\label{eq:Delta_t_most_general}
\end{equation}
where $B_i = b_i$ and $C_i = (c_i - \kappa_i b_i)/\Omega_i$.

Please note, that if we put $c = 0$ in (\ref{eq:delta_s1_app_P_and_Q}), then for any choice of parameters appearing in  $\eta(t)$ (\ref{eq:eta_t}), i.e, $\omega_{\rm L}$, $\gamma$, $k_i$ and $\lambda_i$ we have both $\Delta(0) = 1$ and $\dot{\Delta}(0) = 0$. However, if the those numbers are a priori unknown, and we fit $\Delta(t)$ of the form (\ref{eq:Delta_t_most_general}) to experimental data, both constraints must be enforced.\\

\tocless\section{Extended variable formalism: Derivation of the formulas (29) in the main text}
For the sake of clarity, below we present a step-by-step derivation: First, for a single exponential in the memory kernel, and then for the full model used in the main text.\\

\tocless\subsection{Arbitrary potential, single exponential in the memory kernel}
With an arbitrary potential, $V(z)$, the deterministic force is equal to $- V'(z(t)) \equiv -\frac{\partial}{\partial (z)}V(z(t))$ and the generalized Langevin equation (GLE) 
\be \label{eq:gleSI}
\frac{d^2}{dt^2} z = -\frac{1}{m_L}V'(z(t))  - \int_0^t  \eta(t-\tau) \frac{d z(\tau)}{dt} d\tau + F(t)
\ee
has the memory kernel
\be
\eta(t)=\omega_L^2 \gamma \delta(t) + \omega_L^2 k \exp(-\lambda t),
\ee
such that
\be
\langle F(t+s) F(t) \rangle = \eta(s) \langle v^2 \rangle,
\ee
where
\be
\frac{d}{dt} z(t) = v(t),
\ee
and Eq. \ref{eq:gleSI} can be written as:
\be \label{eq:gle_v}
\frac{d}{dt}v= -\frac{1}{m_L}V'(z(t))  - \int_0^t  \eta(t-\tau) v(\tau) d\tau + F(t)
\ee

We split the stochastic force $F(t)$ into two components,
\be
F(t) = F_0(t) + F_1(t),
\ee
such that their time correlations are delta function and exponential:
\be \label{eq:corr_delta}
\langle F_0(t+s) F_0(t) \rangle = \omega_L^2 \gamma  \langle v^2 \rangle \delta(s),
\ee
\be \label{eq:corr_exp}
\langle F_1(t+s) F_1(t) \rangle = \omega_L^2 k  \langle v^2 \rangle \exp(-\lambda s).
\ee
A random force that obeys Eq. \ref{eq:corr_delta} is Gaussian white noise,
\be \label{eq:F0}
F_0(t) = \sqrt{2 \omega_L^2 \gamma  \langle v^2 \rangle } \ \xi_0(t),
\ee
whereas the Eq. \ref{eq:corr_exp} is fulfilled by Ornstein-Uhlenbeck process,
\be \label{eq:F1}
\frac{d}{dt} F_1(t) = -\lambda F_1 + \sqrt{2 \lambda \omega_L^2 k  \langle v^2 \rangle} \  \xi_1(t).
\ee
$\xi_0$, $\xi_1$ are independent Gaussian white noises: $\langle\xi_0(t+s) \xi_0(t) \rangle = \delta(s)$, $\langle\xi_1(t+s) \xi_1(t) \rangle = \delta(s)$, $\langle\xi_0(t) \xi_1(t) \rangle = 0$.\\

Now Eq. \ref{eq:gle_v} takes on the form:
\be
\frac{d}{dt}v= -\frac{1}{m_L}V'(z(t))  - \int_0^t  \omega_L^2 \gamma \delta(t-\tau) v(\tau) d\tau  - \int_0^t  \omega_L^2 k \exp[-\lambda (t-\tau)] v(\tau) d\tau  + F_1(t) + \sqrt{2 \omega_L^2 \gamma  \langle v^2 \rangle} \ \xi_0(t)
\ee
The second term with the delta function can be readily integrated and it yields $\omega_L^2 \gamma v(t)$. For the third term we use an auxiliary variable:
\be
y(t) = - \int_0^t  \omega_L^2 k \exp[-\lambda (t-\tau)] v(\tau) d\tau .
\ee
We calculate the time derivative of $y(t)$ using the Leibniz integral rule, $\frac{d}{dt} \int_{s_1(t)}^{s_2(t)} f(s,t) ds  =  f(s_2,t)\frac{ds_2}{dt} - f(s_1,t)\frac{ds_1}{dt} + \int_{s_1(t)}^{s_2(t)} \frac{\partial f(s,t)}{\partial t} ds$, where $s_1(t)=const.=0$ and $s_2(t) = t$:
\be \label{eq:y}
\frac{d}{dt}y(t) = - \lambda y(t) - \omega_L^2 k v(t).
\ee
Now we have the following system of equations:
\be
\frac{d}{dt} z(t) = v(t)
\ee
\be 
\frac{d}{dt}y(t) = - \lambda y(t) - \omega_L^2 k v(t) \ \  \mathrm{(from \ Eq. \ (\ref{eq:y}))}
\ee
\be \label{eq:3eqv1}
\frac{d}{dt}v(t) = -\frac{1}{m_L}V'(z(t))  - \omega_L^2 \gamma v(t) + y(t) +F_1(t)+  \sqrt{2 \omega_L^2 \gamma  \langle v^2 \rangle} \ \xi_0(t)
\ee
\be 
\frac{d}{dt} F_1(t) = -\lambda F_1 + \sqrt{2 \lambda \omega_L^2 k  \langle v^2 \rangle} \  \xi_1(t) \ \ \mathrm{(from \  Eq. \ (\ref{eq:F1}))}.
\ee
Now we merge $y(t)$ and $F_1(t)$ into a single new variable, $w(t) = y(t)+F_1(t)$, and, accordingly, $\frac{dw(t)}{dt} = \frac{dy(t)}{dt}+\frac{dF_1(t)}{dt}$. After that substitution, the GLE can be written as a system of three equations:
\be
\frac{d}{dt} z(t) = v(t)
\ee
\be \label{eq:3eqv}
\frac{d}{dt}v(t) =-\frac{1}{m_L}V'(z(t))  - \omega_L^2 \gamma v(t) + w(t) +  \sqrt{2 \omega_L^2 \gamma  \langle v^2 \rangle} \ \xi_0(t)
\ee
\be
\frac{d}{dt} w(t) = -\lambda w(t) - \omega_L^2 k v(t) +  \sqrt{2 \lambda \omega_L^2 k  \langle v^2 \rangle} \ \xi_1(t).
\ee
From the equipartition of energy, we have
\be
 \langle v^2 \rangle = \frac{k_B T}{m_L}.
\ee
The pseudo-mass parameter is given by (see the main text):
\be
m_L = \frac{2 \lambda_s}{\omega_L^2}.
\ee
Thus, $\langle v^2 \rangle ={k_B T \omega_L^2} / ({2 \lambda_s} )$ can be substituted into the Eq. \ref{eq:3eqv}.\\

\tocless\subsection{Arbitrary potential, multiple exponentials in the memory kernel}
If the memory kernel contains multiple exponentials,
\be
\eta(t)=\omega_L^2 \gamma \delta(t) +\omega_L^2 \sum_{i=1}^N  k_i \exp(-\lambda_i t),
\ee
then the stochastic force will be split into $N+1$ components (as in \cite{Baczewski2013}, but here we have one more component for the delta function):
\be
F(t) = F_0(t) + \sum_{i=1}^N F_i(t),
\ee
and analogously to the previous derivation, we get a set of auxiliary variables $w_i(t)$, $i=1..N$, such that the GLE can be presented in the form of a set of equations with Gaussian white noises $\xi_i(t)$:
\begin{align} \label{eq:gleN}
\frac{d}{dt} z(t) &= v(t)\\
\frac{d}{dt}v(t) &= - \frac{1}{m_L} V'(z(t)) - \omega_L^2 \gamma v(t) + \sum_{i=1}^N w_i(t) +  \omega_L^2\sqrt{\frac{\gamma k_B T }{ \lambda_s}} \ \xi_0(t)\\
\frac{d}{dt} w_1(t) &= -\lambda_1 w_1(t) - \omega_L^2 k_1 v(t) + \omega_L^2 \sqrt{ \frac{\lambda_1 k_1 k_B T }{ \lambda_s} } \ \xi_1(t)\\
&...\\
\frac{d}{dt} w_N(t) &= -\lambda_N w_N(t) - \omega_L^2 k_N v(t) + \omega_L^2 \sqrt{ \frac{\lambda_N k_N k_B T }{ \lambda_s} } \ \xi_N(t)
\end{align}
\tocless\section{Euler-Maruyama integration scheme}
The simplest numerical integration scheme is the Euler-Maruyama scheme \cite{Mannella2002}. The increment of the deterministic part is $\Delta t$, and the increment of the stochastic part is $\sqrt{\Delta t}$. An intuitive explanation for this square-root dependence on time is the following: The probability distribution of distances travelled by a Wiener process $W_i$ (a Brownian walk trajectory whose evolution is described by $dW_i = \xi_i(t)dt$),  is Gaussian with the variance growing proportionally to time. And therefore, standard deviation (a quantity of the dimension of distance increment) is proportional to the square root of time. Then, if we draw a number from a normal distribution centered at 0, whose variance is equal to $1$, and then multiply it by $\sqrt{A} \sqrt{\Delta t}$ ($A$ being some number), this is equivalent to rescaling of that normal distribution, i.e. drawing a number from a normal distribution centered at 0, whose variance is $A \Delta t$.

The numerical algorithm is as follows:
\begin{align} \label{eq:gleNnum}
z(t+\Delta t)  &=z(t) + v(t) \Delta t\\
v(t+\Delta t)   &=v(t) + \left[- \frac{1}{m_L}V'(z(t)) - \omega_L^2 \gamma v(t)  + \sum_{i=1}^N w_i(t)\right]{\Delta t}  +   \left[ \omega_L^2 \sqrt{\frac{\gamma k_B T }{ \lambda_s}}\right]\sqrt{\Delta t}\ \zeta_0\\
w_1(t+\Delta t) &=  w_1(t)+ \left[-\lambda_1 w_1(t) - \omega_L^2 k_1 v(t)\right]{\Delta t}  + 
\left[ \omega_L^2 \sqrt{\frac{\lambda_1 k_1 k_B T }{ \lambda_s}} \right] \sqrt{\Delta t}\ \zeta_1\\
&...\\
w_N (t+\Delta t)  &= w_N(t) + \left[-\lambda_N w_N(t) - \omega_L^2 k_N v(t)\right]{\Delta t}  +  \left[ \omega_L^2 \sqrt{\frac{\lambda_N k_N k_B T }{ \lambda_s}} \right] \sqrt{\Delta t} \ \zeta_N 
\end{align}
Here, $\zeta_i$ are numbers drawn separately from the normal distribution of the mean $0$ and variance $1$.\\

\tocless\section{Details on Friction}
The changes in the electric field of the medium in the solute's cavity can be traced back to the material diffusive motion of the dipoles of the solvent molecules. In fact, the linear solvent polarization mass, $m_{\rm L}$,\footnote{Be aware, that this mass is not identical for different solutes as it depends on the reorganization energy and thus on the size of the cavity in which the solute resides.} can be related to the solvent molecular moment of inertia, $I$ (strictly valid for linear and spherical top molecules).\cite{hynes_JPC_1986}
\begin{align}
m_{\rm L} & = \dfrac{2\lambda_{\rm s}}{\omega_{\rm L}^2} \label{eq:mL}\\
\omega_{\rm L}^2 & = \omega_{\rm f}^2 \left(\dfrac{2\epsilon_0 + \epsilon_\infty}{3\epsilon_\infty g_{\rm K}} \right)\\
\omega_{\rm f}^2 & = 2\dfrac{k_{\rm B}T}{I}
\end{align}
$\lambda_{\rm s}$ is the solvent reorganization energy, $\epsilon_0$ and $\epsilon_\infty$ are the dielectric constants at zero (static) and infinite frequency (optical) respectively, $\omega_{\rm L}$ and $\omega_{\rm f}$ are the longitudinal and free rotational frequencies of the solvent respectively, and $g_{\rm K}$ is the Kirkwood correlation factor (1 for solvent molecules that are not correlated, and larger than 1 for solvent molecules that are strongly interacting with each other creating local structures). The inertial moment of the solvent molecule can be calculated from the rotational constant obtained by means of microwave spectroscopy:
\begin{equation}
B = \dfrac{h}{4\pi c I},
\end{equation}
with $B$ given in wavenumbers. The rotational constants for almost any molecule can be found in ref.\ \citenum{nist_rot}.\footnote{Any non-spherical molecule will possess more than one rotational moment. Weaver\cite{mcmanis_CP_1991}  and Fawcett\cite{fawcett_2004} have taken the largest moment. Remember that this moment is equal to the mass of the rigid rotor times the square of its gyration radius, so for the largest moment calculated for ACN, the corresponding gyration radius is only \unit[1.15]{\angstrom}. Taking the other smaller moment will translate into an even smaller radius.} We also need the Kirkwood correlation factor, which in turn can be calculated if not found experimentally with the Mean Spherical Approximation (MSA) model expression:
\begin{equation}
g_{\rm K} = \dfrac{(\epsilon - 1)(2\epsilon + 1)}{3\epsilon d_{\rm p}^2}
\end{equation}
in which the dielectric constant is $\epsilon$ and $d_{\rm p}$ is the dipole polarizability. This latter quantity is obtained by recursively solving the equations associated to it in the MSA using just the solvent properties of dipole moment, density, radius and polarizability (see ref.\ \citenum{blum_JPC_1993}).\\

\tocless\section{Additional Experimental Data}

\begin{figure}[h!]
\centering
	\includegraphics[scale=1.25]{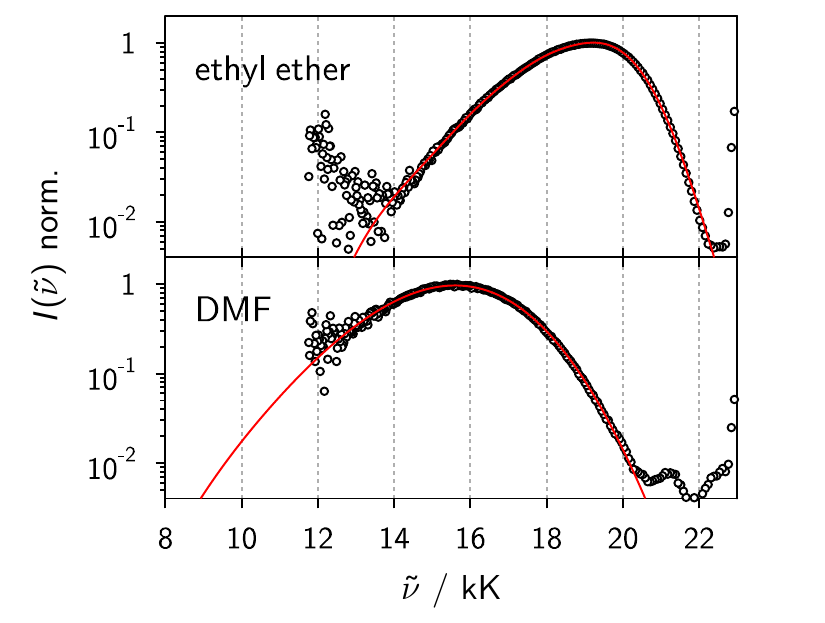}
\caption[Lineshape fits to steady-state emission]{Exemplary fits of the lineshape function convolved with a Gaussian (red line, using eqs.~7 to 9 in main manuscript) to steady-state emission spectra (open circles) for ethyl ether (6) and dimethylformamide (13).}
\label{fig:}
\end{figure}

\begin{figure}[h!]
\centering
	\includegraphics[scale=1.25]{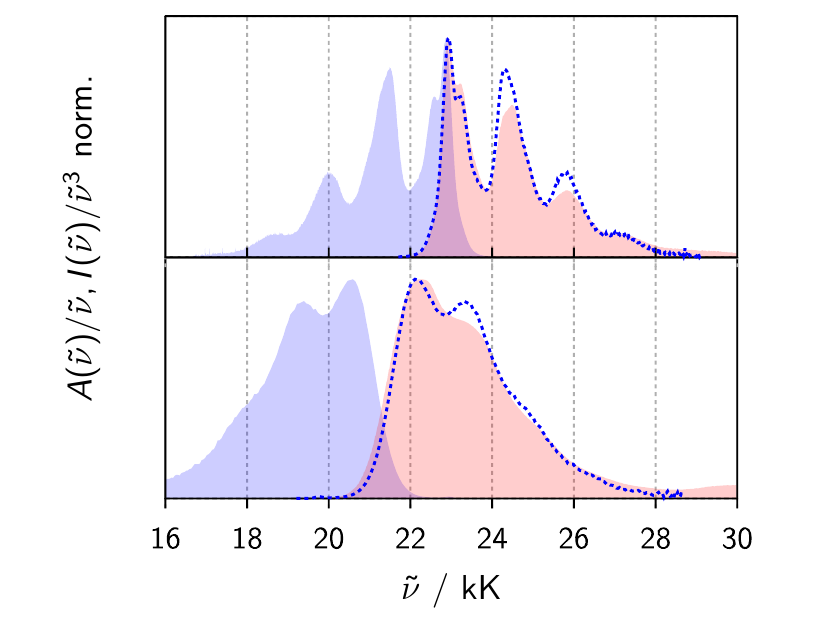}
\caption[Mirror symmetry for Pe and PeDMA]{Comparison of the mirror symmetry for \Pe\ (above) and \PD\ (below) in cyclohexane. The dashed blue line shows the mirrored fluorescence, $I(2\tnu_{00} - \tnu)/\tnu^3$, with $\tnu_{00}$ being the 00 transition wavenumber.}
\label{fig:}
\end{figure}

\begin{figure}[p]
\centering
	\includegraphics[scale=1.25]{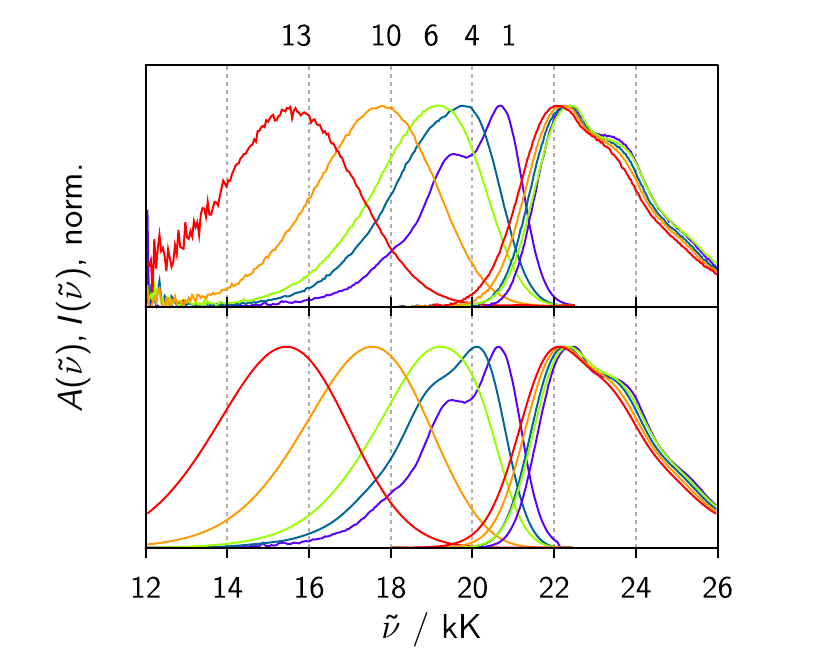}
\caption[Simulated vs.\ experimental steady-state spectra]{Comparison of experimental (upper panel) and simulated (lower panel) steady-state spectra (numbers above the graphic refer to the solvent number in Table~1 of the manuscript, see section IV. A in main manuscript). Beware, that the increasing noise in the red part of the emission spectra is owed to the decreasing quantum efficiency of the detector.}
\label{fig:}
\end{figure}

\begin{figure}[p]
\centering
	\includegraphics[scale=1.25]{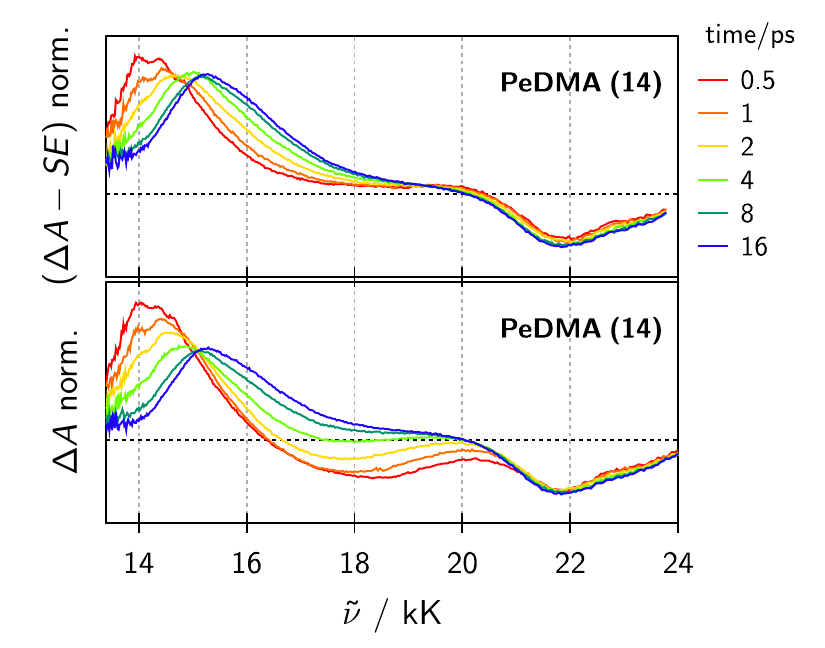}
\caption[fs-TA in dimethylsulfoxide, corrected for stimulated emission]{fs-TA spectra (lower panel) and fs-TA spectra with subtracted stimulated emission (upper panel) of \PD\ in dimethylsulfoxide (14). FLUPS data were transferred to stimulated emission using $I_{\rm SE} = I(\tnu)/\tnu^2$.}
\label{fig:}
\end{figure}

\begin{figure}[p]
\centering
	\includegraphics{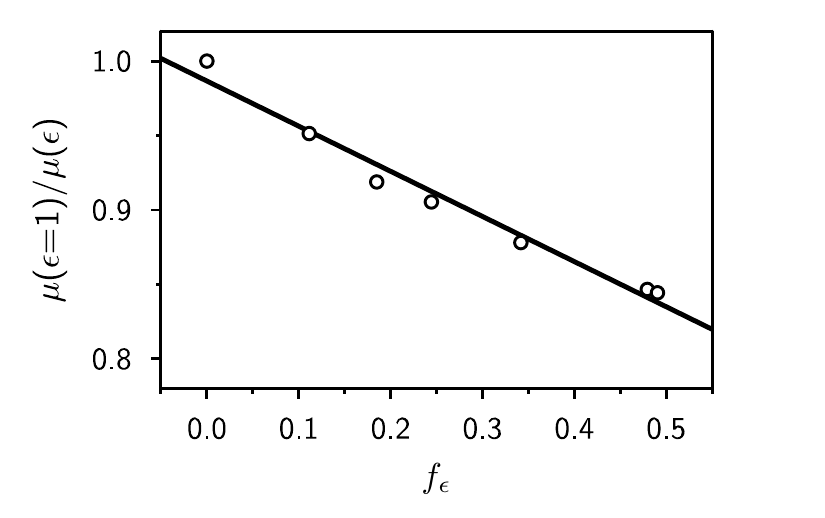}
\caption[Calculated solvent dependence of the dipole moment of PeDMA]{Solvent dependence of the calculated ground state dipole moment of \PD\ obtained from QM calculations.}
\label{fig:}
\end{figure}

\begin{figure}[p]
	\includegraphics[scale = 1.25]{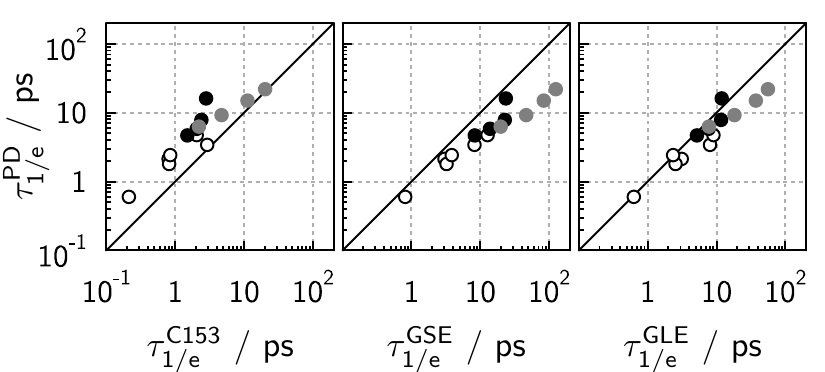}
\caption[Comparison of phenomenological time-scales between C153 and PeDMA]{Comparison of $1/e$-times, $\tau_{1/e}$, for the dynamic Stokes shift of \PD\ with the corresponding times for \C\ and using simulations with the GSE and the GLE. White circles ($\circ$) are pure solvents, full black circles ($\bullet$) denote isoviscous binary mixtures and grey circles (\textcolor{lightgray}{$\bullet$}) denote isodelectric binary mixtures.}
\label{fig:taus}
\end{figure}

\begin{figure}[p]
	\includegraphics[]{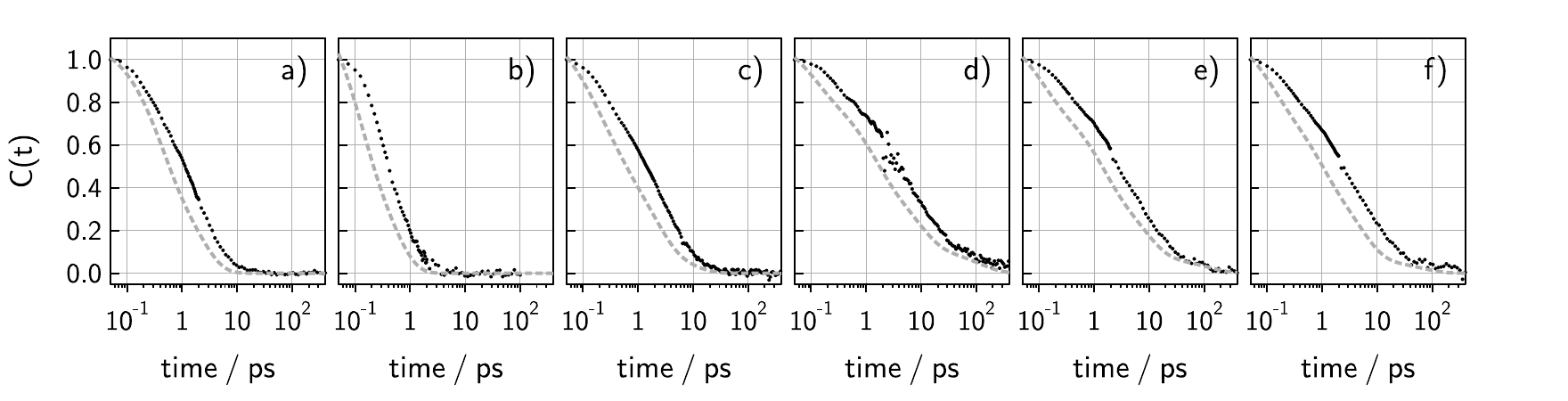}\\
	\includegraphics[]{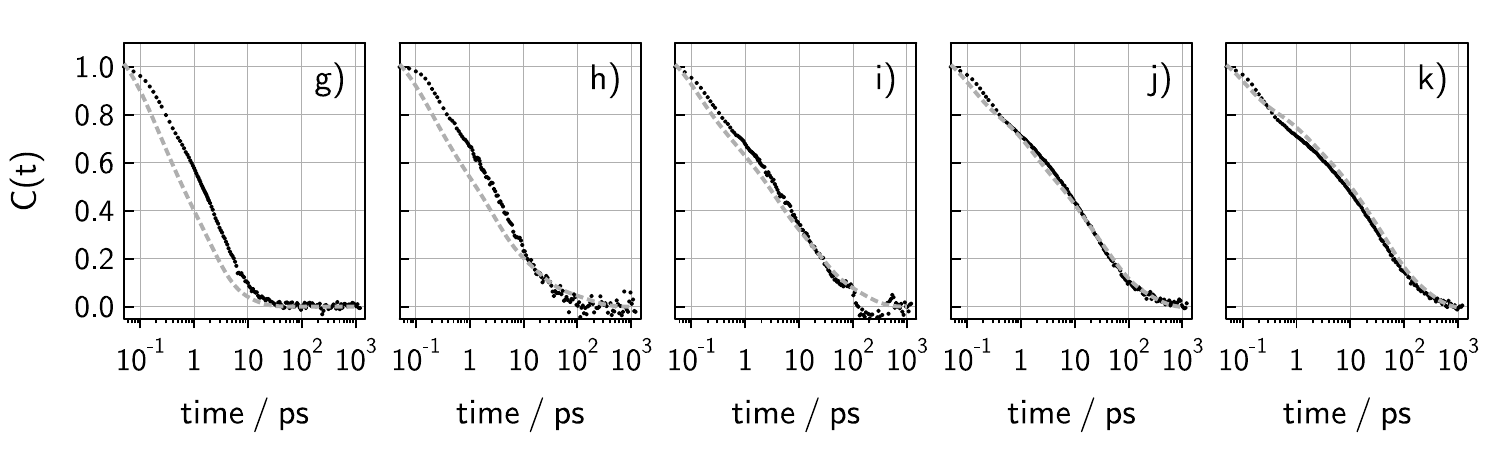}
\caption[Comparison of $C(t)$ for PeDMA and C153 in selected polar solvents]{Comparison of $C(t)$ for \PD\ (black dots) and \C\ (grey dashed lines) in polar solvents. a) - c) is butyronitrile, acetonitrile and dimethylsulfoxide. d) - f) are binary mixtures of DB with $\epsilon$ of 16 (13), 23 (20) and 30 (31). g) - k) are binary mixtures of DG with $\eta$ of 2, 5, 10, 20 and \unit[40]{cP}.}
\label{fig:taus}
\end{figure}

\begin{figure}[p]
	\includegraphics{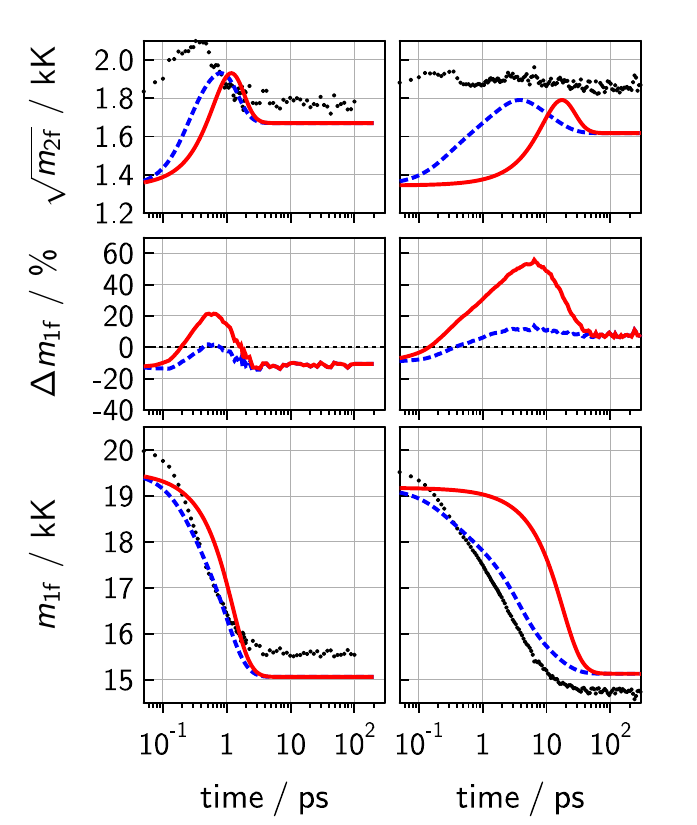}
\caption[Comparison of GSE and Smoluchowski equation with $D_0$ for two solvents]{Comparison of experimental (black dots) time dependence of the spectral moments with simulations using the GSE (with entire $D(t)$, blue dashed line) and the Smoluchowski equation using $D_0$ (i.\,e.\ the long time diffusion coefficient, red full line) in acetonitrile (left) and dimethylsulfoxide (right).}
\label{fig:}
\end{figure}

\begin{figure}[p]
      \includegraphics[width=0.49\textwidth]{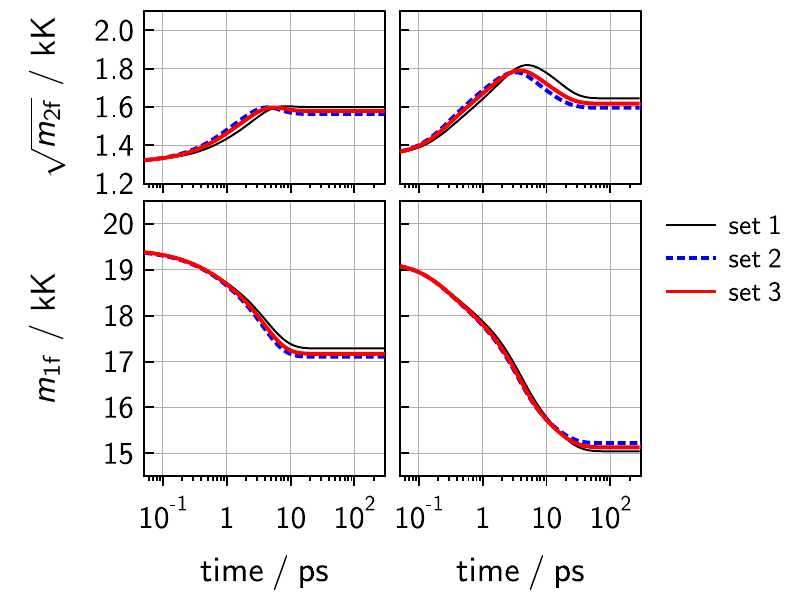}
    \hfill
      \includegraphics[width=0.49\textwidth]{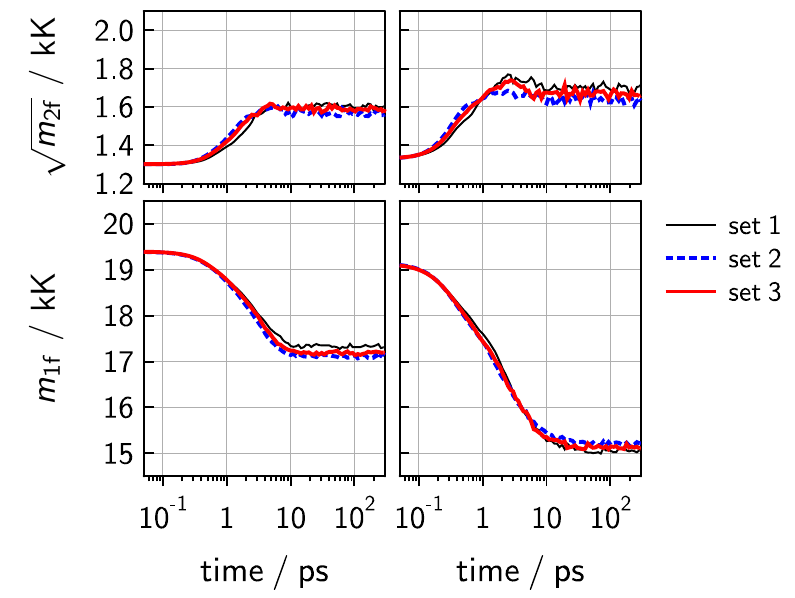}
\caption[Comparison of simulations with 3 different 3-level-model parameter-sets]{Comparison of simulation results using 3 different parameter sets for the FES of \PD\ using the GSE (left) and GLE (right) in tetrahydrofuran and dimethylsulfoxide.\\
set 1: $U_{\rm l}/hc = 22.887$\,kK; $U_{\rm c}/hc = 27.455$\,kK; $J/hc = 1.114$\,kK; $\mu_{\rm l} = 9.33$\,D; $\mu_{\rm c} = 33.96$\,D; $2D/hca^3 = 4.294$\,kK\\
set 2: $U_{\rm l}/hc = 22.774$\,kK; $U_{\rm c}/hc = 26.419$\,kK; $J/hc = 1.256$\,kK; $\mu_{\rm l} = 4.19$\,D; $\mu_{\rm c} = 32.19$\,D; $2D/hca^3 = 3.331$\,kK\\
set 3: $U_{\rm l}/hc = 22.898$\,kK; $U_{\rm c}/hc = 27.024$\,kK; $J/hc = 1.325$\,kK; $\mu_{\rm l} = 5.80$\,D; $\mu_{\rm c} = 33.25$\,D; $2D/hca^3 = 3.887$\,kK}
\label{fig:}
\end{figure}

\begin{figure}[h]
	\includegraphics[scale=1.5]{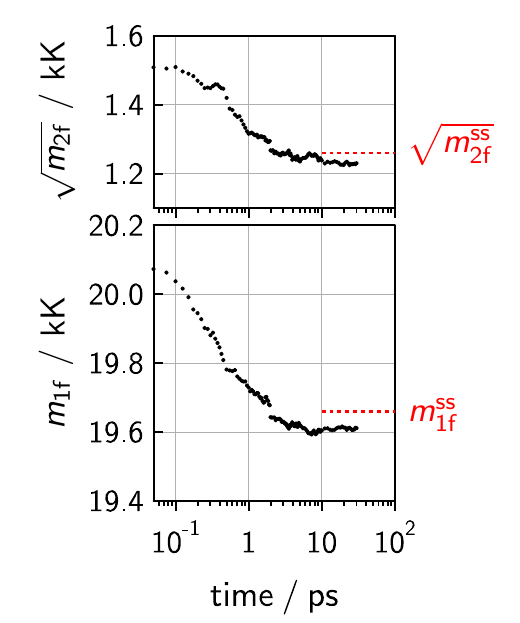}
\caption[Time dependence of spectral moments for PeDMA in cyclohexane]{Time-dependence of mean fluorescence frequency, $m_{\rm 1f}$, and spectral standard deviation, $\sqrt{m_{\rm 2f}}$, of \PD\ in cyclohexane (upon excitation at \unit[400]{nm}).}
\label{fig:}
\end{figure}

\setlength{\tabcolsep}{20pt}
\begin{table*}[p]
\begin{threeparttable}[c]
     	\caption{\C-parameters establishing the FES in solvents of arbitrary $\epsilon$ and $n$.}
	\label{tab:pars}
		\begin{tabular}{c..}
		\toprule
			& \multicolumn{2}{c}{state-specific} \\
			 			& \multicolumn{1}{c}{$i = {\rm g}$} & \multicolumn{1}{c}{$i = {\rm c}$}\\
		\midrule
		 $\mu_i$ / D		& 6.8 	& 15.8 \\
		$\mu_{i{\rm g}}$ / D 	& 	& 6 \\
		$U_i/hc$ / kK	& 0   	& 25.35 \\[1ex]
			& \multicolumn{2}{c}{common}\\
		\cline{2-3}
		\rule{0pt}{4ex}
		$J/hc$ / kK 		& 	 \multicolumn{2}{c}{1.7}	 \\
		$\alpha / a^3$   	&   	 \multicolumn{2}{c}{0.5}	\\
		$a$ / \AA			&  	 \multicolumn{2}{c}{6.5}	\\
		$2D/hca^3$ / kK	&	 \multicolumn{2}{c}{5.2} 	\\
		\bottomrule	
		\end{tabular}
\end{threeparttable}
\end{table*}

\setlength{\tabcolsep}{5pt}
\begin{table}[p]
\begin{threeparttable}
     	\caption[Parameters for the solvent shift dynamics of C153]{Parameters for the solvent shift dynamics of \C, obtained from fitting eq.~11 of the manuscript to the peak frequencies of the log-normal fits of the time-resolved emission spectra of \C.}
	\label{tab:pars}
		\begin{tabular}{l...........}
		\toprule
		solvent & \multicolumn{1}{c}{$\Delta\tnu_{\rm theo}$} & \multicolumn{1}{c}{$\Delta\tnu_1$} & \multicolumn{1}{c}{$\tau_1$\tnote{b}} & \multicolumn{1}{c}{$\Delta\tnu_2$} & \multicolumn{1}{c}{$\tau_2$} & \multicolumn{1}{c}{$\Delta\tnu_3$} & \multicolumn{1}{c}{$\tau_3$} & \multicolumn{1}{c}{$\Delta\tnu_4$} & \multicolumn{1}{c}{$\tau_4$} & \multicolumn{1}{c}{$\Delta\tnu_5$} & \multicolumn{1}{c}{$\tau_5$} \\
			& \multicolumn{1}{c}{(kK)} & \multicolumn{1}{c}{(kK)} & \multicolumn{1}{c}{(ps)} & \multicolumn{1}{c}{(kK)} & \multicolumn{1}{c}{(ps)} & \multicolumn{1}{c}{(kK)} & \multicolumn{1}{c}{(ps)} & \multicolumn{1}{c}{(kK)} & \multicolumn{1}{c}{(ps)} & \multicolumn{1}{c}{(kK)} & \multicolumn{1}{c}{(ps)} \\
		\midrule
		{\it i}-propyl ether (6) & 0.71 & 0.02 & 0.10 & 0.20 & 0.35 & 0.44 & 4.1 & 0.05 & 50 & - & -   \\ 
		butyl acetate (8) & 1.20 & 0.19 & 0.10 & 0.32 & 0.49 & 0.49 & 3.2 & 0.19 & 24 & - & -   \\ 
		tetrahydrofuran (10) & 1.22 & 0.09 & 0.10 & 0.48 & 0.36 & 0.65 & 1.6 & - & - & - & - \\ 
		butyronitrile (11) & 1.48 & 0.00 & 0.10 & 0.80 & 0.36 & 0.71 & 2.0 & - & - & - & -  \\ 
		acetonitrile (12) & 1.74 & 0.90 & 0.10 & 0.84 & 0.47 & - & -  & - & - & - & -  \\ 
		dimethylsulfoxide (14) & 1.84 & 0.38 & 0.10 & 0.53 & 0.30 & 0.73 & 1.9 & 0.20 & 8.3 & - & - \\[1ex]
		
		DB($\epsilon=6$)\tnote{a} & 1.17 & 0.10 & 0.10 & 0.37 & 0.37 & 0.47 & 4.1 & 0.23 & 33 & - & - \\ 
		DB($\epsilon=9$) & 1.42 & 0.38 & 0.10 & 0.41 & 1.3 & 0.45 & 9.0 & 0.18 & 160 & - & -  \\ 
		DB($\epsilon=13$) & 1.53 & 0.36 & 0.10 & 0.52 & 1.2 & 0.48 & 8.3 & 0.17 & 110 & - & -  \\ 
		DB($\epsilon=20$) & 1.62 & 0.47 & 0.10 & 0.54 & 1.2 & 0.50 & 8.0 & 0.11 & 120 & - & -  \\ 
		DB($\epsilon=31$) & 1.72 & 0.46 & 0.10 & 0.52 & 0.73 & 0.63 & 4.8 & 0.10 & 62 & - & -  \\[1ex]
		
		DG(\unit[5]{cP})\tnote{b} & 1.97 & 0.35 & 0.10 & 0.46 & 0.30 & 0.63 & 2.3 & 0.38 & 15 & 0.16 & 140   \\ 
		DG(\unit[10]{cP}) & 2.07 & 0.47 & 0.10 & 0.30 & 0.40 & 0.52 & 2.8 & 0.51 & 21 & 0.27 & 150   \\ 
		DG(\unit[20]{cP}) & 2.10 & 0.48 & 0.10 & 0.42 & 1.2 & 0.36 & 7.8 & 0.56 & 41 & 0.27 & 230   \\ 
		DG(\unit[40]{cP}) & 2.13 & 0.47 & 0.10 & 0.30 & 1.2 & 0.34 & 7.7 & 0.63 & 44 & 0.39 & 250  	\\	
		\bottomrule
		\end{tabular}
\begin{tablenotes}		
		\item [a] DB - binary mixtures of dimethylsulfoxide / benzyl acetate
		\item [b] DG - binary mixtures of dimethylsulfoxide / glycerol
	\end{tablenotes}
\end{threeparttable}
\end{table}

\setlength{\tabcolsep}{5pt}
\begin{table}[p]
     	\caption[Parameters for calculating the time-dependent friction $\eta(t)$]{Parameters for calculating the time-dependent friction $\eta(t)$ (see eqs.~1 and 28 in the main manuscript and eq.\,\eqref{eq:mL})}
	\label{tab:pars}
		\begin{tabular}{@{}l........c.@{}}
		\toprule
		solvent &  \multicolumn{1}{c}{$m_{\rm L}$} & \multicolumn{1}{c}{$\lambda_{\rm s}$} & \multicolumn{1}{c}{$\omega_{\rm L}^2$} & \multicolumn{1}{c}{$\gamma$} & \multicolumn{1}{c}{$\omega_{\rm L}^2 k_1$} & \multicolumn{1}{c}{$1/\lambda_1$} & \multicolumn{1}{c}{$\omega_{\rm L}^2 k_2$} & \multicolumn{1}{c}{$1/\lambda_2$} & \multicolumn{1}{c}{$\omega_{\rm L}^2 k_3$} & \multicolumn{1}{c}{$1/\lambda_3$} \\ 
			&	 \multicolumn{1}{c}{(eVps$^2$)} & \multicolumn{1}{c}{(eV)} & \multicolumn{1}{c}{(ps$^{-2}$)} & \multicolumn{1}{c}{(ps)} & \multicolumn{1}{c}{(ps$^{-2}$)} & \multicolumn{1}{c}{(ps)} & \multicolumn{1}{c}{(ps$^{-2}$)} & \multicolumn{1}{c}{(ps)} & \multicolumn{1}{c}{(ps$^{-2}$)} & \multicolumn{1}{c}{(ps)}\\
\midrule
				{\it i}-propyl ether (6)  & 0.3043 & 0.171 & 1.13 & 0.75 & 5.36 & 0.043 & 1.61 & 1.51 & 7.46 10$^{-2}$ & 45.8 \\ 
				butyl acetate (8)  & 0.2979 & 0.205 & 1.37 & 0.3 & 2.91 & 0.158 & 1.10 & 1.73 & 2.17 10$^{-1}$ & 20.2 \\ 
				tetrahydrofuran (10)  & 0.0691 & 0.255 & 7.39 & 0.6 & 8.45 & 0.102 & 3.62 & 0.93 & 4.47 10$^{-4}$ & 9.8 \\ 
				butyronitrile (11)  & 0.0402 & 0.349 & 17.38 & 0.6 & 22.35 & 0.011 & 7.68 & 1.20 & 6.59 10$^{-4}$ & 15.1 \\ 
				acetonitrile (12)  & 0.0073 & 0.378 & 103.23 & 0.15 & 26.06 & 0.010 & 41.81 & 0.29 & 1.82 10$^{-3}$ & 6.2 \\ 
				dimethylsulfoxide (14)  & 0.0194 & 0.332 & 34.28 & 0.3 & 21.60 & 0.162 & 20.29 & 1.14 & 3.04 & 7.5 \\[1ex] 

				DB($\epsilon=13$)  & 0.1439 & 0.272 & 3.78 & 0.5 & 7.22 & 0.511 & 1.73 & 5.60 & 4.49 10$^{-1}$ & 99.9 \\ 
				DB($\epsilon=20$)  & 0.0545 & 0.241 & 8.83 & 0.5 & 12.46 & 0.580 & 3.52 & 5.52 & 6.18 10$^{-1}$ & 107.1 \\ 
				DB($\epsilon=31$)  & 0.0346 & 0.318 & 18.40 & 0.5 & 22.38 & 0.363 & 9.22 & 3.09 & 1.10 & 58.3 \\[1ex]

				DG(\unit[5]{cP})  & 0.0198 & 0.334 & 33.78 & 0.5 & 35.04 & 0.862 & 9.14 & 6.40 & 5.02 & 51.6 \\ 
				DG(\unit[10]{cP})  & 0.0202 & 0.335 & 33.09 & 0.5 & 41.08 & 1.181 & 13.78 & 14.47 & 4.49 & 125.7 \\ 
				DG(\unit[20]{cP})  & 0.0211 & 0.335 & 31.73 & 0.5 & 52.84 & 1.132 & 19.76 & 17.13 & 6.05 & 137.6 \\ 
				DG(\unit[40]{cP})  & 0.0219 & 0.334 & 30.56 & 0.5 & 50.30 & 1.639 & 23.50 & 19.41 & 7.70 & 150.1 \\
				\bottomrule
		\end{tabular}
\end{table}

\clearpage
\tocless%

\end{document}